%% file: main.tex
\pdfoutput=1

\documentclass[11pt]{article}

\usepackage[final]{acl}

\usepackage{times}
\usepackage{latexsym}

\usepackage[T1]{fontenc}

\usepackage[utf8]{inputenc}

\usepackage{microtype}

\usepackage{inconsolata}

\usepackage{graphicx}

\usepackage{tikz}
\usepackage{amsmath}

\usepackage{filecontents}
\usepackage{xurl,MnSymbol}

\usepackage{xcolor}
\usepackage{wasysym}
\usepackage{fontawesome}
\usepackage{ifthen}
\usepackage[table]{xcolor}

\usepackage{colortbl}
\newboolean{textswitch}
\setboolean{textswitch}{false} 

\input{prestuff.tex}

\usepackage{enumitem} 
\begin{document}

\newcommand{\system}{{\sc WaterPark}\xspace}
\newcommand{\ting}[1]{\textcolor{red}{[#1]}\xspace}
\newcommand{\jc}[1]{\textcolor{blue}{[#1]}\xspace}
\newcommand{\lauren}[1]{{\textcolor{purple}{[#1]}}\xspace}
\newcommand{\jiang}[1]{{\textcolor{violet}{[#1]}}\xspace}
\newcommand{\zian}[1]{{\textcolor{magenta}{[#1]}}\xspace}



\date{}

\title{\includegraphics[height=1.2em]{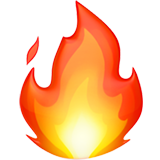} Watermark under Fire: A Robustness Evaluation of LLM Watermarking}
\author{
\textbf{Jiacheng Liang$^{1}$, 
Zian Wang$^{1}$, 
Spencer Hong$^{2}$, 
Shouling Ji$^{3}$, 
Ting Wang$^{1}$} \\ 
$^1$Stony Brook University \,\,
$^2$National University of Singapore  \,\,
$^3$Zhejiang University \\
\texttt{ljcpro@outlook.com}, 
\texttt{twang@cs.stonybrook.edu}
}

\maketitle

\input{abstract}

\input{introduction}

\input{taxonomy}

\input{platform}
\input{eval}

\input{attack}
\input{discussion}

\input{conclusion}
\bibliography{bibs/watermark}
\input{appendix}
\end{document}

%% file: prestuff.tex
\usepackage{epsfig,amsmath,amsfonts,epsfig,multirow,makecell,caption,soul,csquotes,color,wrapfig,subcaption,mathtools,bm,spverbatim,booktabs,tcolorbox,diagbox}
\usepackage[e]{esvect}

\makeatletter
\newif\if@restonecol
\makeatother

\usepackage[boxed, ruled, vlined, linesnumbered]{algorithm2e}
\SetKwRepeat{Do}{do}{while}

\newenvironment{changemargin}[2]{\begin{list}{}{
	\setlength{\topsep}{0pt}\setlength{\leftmargin}{0pt}
	\setlength{\rightmargin}{0pt}
	\setlength{\listparindent}{\parindent}
	\setlength{\itemindent}{\parindent}
	\setlength{\parsep}{0pt plus 1pt}
	\addtolength{\leftmargin}{#1}\addtolength{\rightmargin}{#2}
	}\item}
	{\end{list}}

\usepackage[first=0,last=9]{lcg}
\usepackage{colortbl}
\definecolor{Gray}{gray}{0.7}
\colorlet{Red}{red!20!white}
\colorlet{Blue}{blue!20!white}

\usepackage{hyperref}

\newcommand{\msec}[1]{\S\ref{#1}}
\newcommand{\mref}[1]{\,\ref{#1}}
\newcommand{\meq}[1]{Eq.\,\ref{#1}}
\newcommand{\mcite}[1]{\cite{#1}}

\newcommand{\mie}{\textit{i.e.}\xspace}

\newtcolorbox{mtbox}[1]{left=0.25mm, right=0.25mm, top=0.25mm, bottom=0.25mm, sharp corners, colframe=red!50!black, boxrule=0.5pt, title={#1}, fonttitle=\bfseries, coltitle=red!50!black, attach title to upper={\ --\ }}

\usepackage{scalerel}[2016/12/29]

\makeatletter
\providecommand{\leadsfrom}{%
  \mathrel{\mathpalette\reflect@squig\relax}%
}
\newcommand{\reflect@squig}[2]{%
  \reflectbox{$\m@th#1\leadsto$}%
}
\makeatother

\def\eqref#1{equation~\ref{#1}}

\def\1{\bm{1}}

\DeclareMathAlphabet{\mathsfit}{\encodingdefault}{\sfdefault}{m}{sl}
\SetMathAlphabet{\mathsfit}{bold}{\encodingdefault}{\sfdefault}{bx}{n}



%% file: abstract.tex
\begin{abstract}

Various watermarking methods (``watermarkers'') have been proposed to identify LLM-generated texts; yet, due to the lack of unified evaluation platforms, many critical questions remain
under-explored: i) What are the strengths/limitations of various watermarkers, especially their attack robustness? ii) How do various design choices impact their robustness? iii) How to optimally operate watermarkers in adversarial environments?
To fill this gap, we systematize existing LLM watermarkers and watermark removal attacks, mapping out their design spaces. We then develop \system, a unified platform that integrates 10 state-of-the-art watermarkers and 12 representative attacks. More importantly, by leveraging \system, we conduct a comprehensive assessment of existing watermarkers, unveiling the impact of various design choices on their attack robustness. We further explore the best practices to operate watermarkers in adversarial environments. We believe our study sheds light on current LLM watermarking techniques while \system serves as a valuable testbed to facilitate future research.\footnote{All the source code and data are publicly available: \url{https://github.com/JACKPURCELL/WaterPark}}
\end{abstract}


%% file: introduction.tex
\section{Introduction}

The recent advances in large language models (LLMs), including GPT~\citep{chatgpt} and Llama~\citep{llama}, have significantly enhanced our capabilities of general-purpose text generation and complex problem-solving, but also raised concerns about misuse through disinformation~\citep{izacard2022atlas}, phishing~\citep{10.1145/3442188.3445922}, and academic dishonesty~\citep{academic-gpt}. There is thus a pressing need for the capability of identifying LLM-generated content. 

A simple approach is to train classifiers to distinguish between LLM- and human-generated texts~\citep{gptzero}. However, as LLMs improve, this distinction becomes less clear. Watermarking has emerged as an alternative solution, embedding statistical signals (``watermarks'') during generation to verify LLM-produced texts. Various watermarking methods (``watermarkers") have been developed~\citep{scott,aiweia,aiweib,john,xiaoniu,xuandong,rohith}, each with unique design choices and desirable properties, raising a set of intriguing questions:

\begin{figure}[t!]
    \centering
    \includegraphics[width=0.9\linewidth]{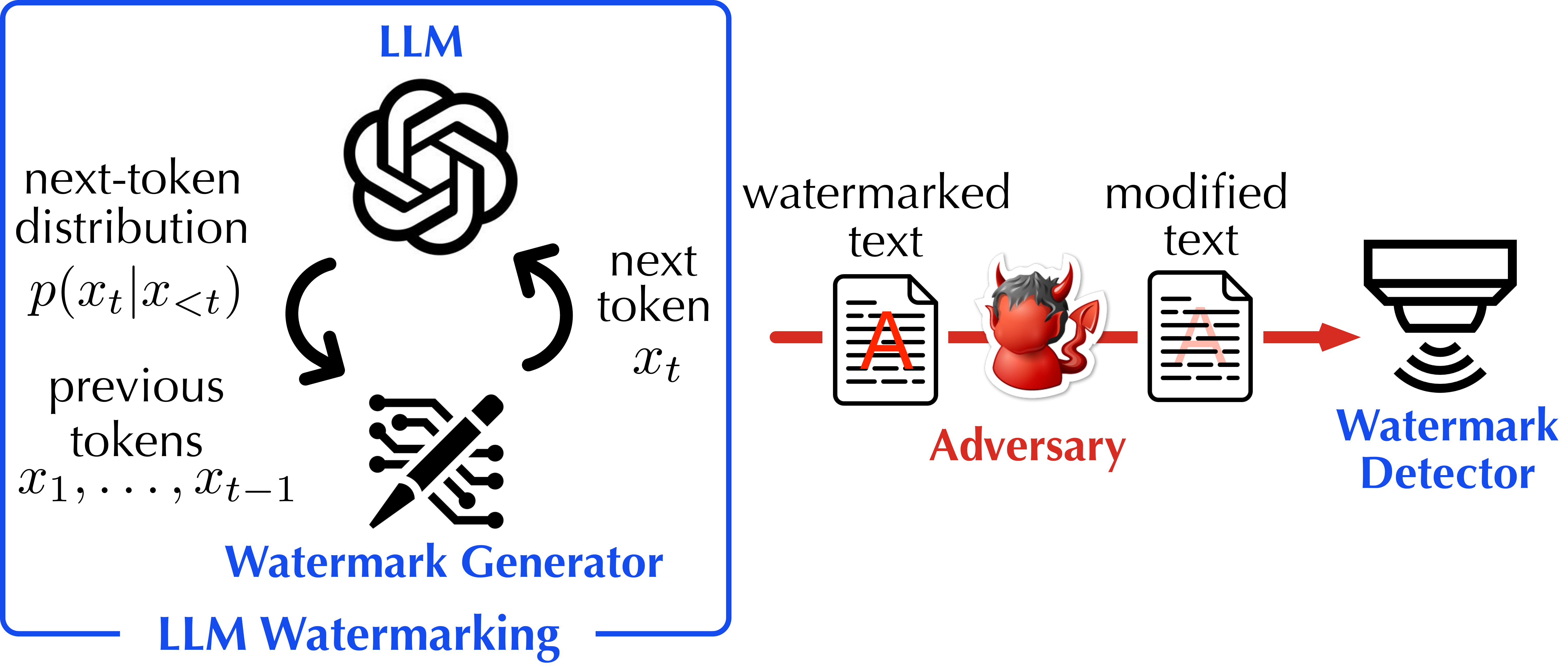}
    \caption{\small Illustration of LLM watermarking and watermark removal attacks.
    \label{fig:flow}}
\end{figure}

RQ1 -- What are the strengths and limitations of various watermarkers, especially their robustness against manipulations?

RQ2 -- How do different design choices impact the attack robustness of watermarkers?

RQ3 -- What are the best practices for operating watermarkers in adversarial environments?

\begin{table*}[!ht]\small
\centering
\renewcommand{\arraystretch}{0.7}
\setlength{\tabcolsep}{2pt}

\vspace{-5pt}
\caption{\small Conclusions in prior work and \system ($\Circle$ -- inconsistent; $\LEFTcircle$ -- partially inconsistent; $\CIRCLE$ -- consistent).}
\resizebox{0.98\textwidth}{!}{
    \begin{tabular}{m{4.5cm}m{5cm}m{4.5cm}c}
    \toprule
    Previous Conclusion &Refined Conclusion  & Explanation & Consist.  \\
        \midrule

UG~\citep{xuandong} is not robust due to its context-free design~\citep{markmyword}. &  UG shows higher resilience than other watermarkers to paraphrasing attacks. & UG's context-free design ensures consistency between detection and generation, avoiding issues in text-dependent designs. & $\Circle$ \\ 
    \midrule
UPV~\citep{aiweia} has a fairly low false positive rate. & UPV is more prone to false positive cases compared to TGRL. & \multirow{2}{4.5cm}{While model-based detection incurs higher uncertainty compared to score-based detection, it fails to offer higher flexibility in countering paraphrasing attacks.}& $\LEFTcircle$  \\ 
\cmidrule{1-2} 
\cmidrule{4-4} 
UPV shows strong robustness to rewriting and outperforms TGRL under paraphrasing attacks. & Both UPV and TGRL struggle against GPT-based paraphrasing attacks. & &$\LEFTcircle$  \\ 
    \midrule
RDF~\citep{rohith} significantly outperforms TGRL against substitution attacks. & RDF shows strong robustness against synonym substitution and other lexical editing attacks. & RDF's distribution-transform strategy is more robust than the distribution-shift strategy against lexical editing. &$\CIRCLE$ \\ 
    \midrule

RDF's~\citep{rohith} edit score-based detection is insensitive to local misalignment caused by token insertion~\citep{markmyword}. & RDF (edit score) shows higher resilience against lexical editing attacks compared to GO (plain score) when token length is fixed. However, this advantage diminishes as token length varies. &The edit score-based detection is robust to lexical editing but sensitive to varying token length. & $\LEFTcircle$  \\ 
    \midrule
SIR~\citep{aiweib} shows strong resilience against paraphrasing attacks. &SIR's effectiveness decreases as the intensity of paraphrasing attacks increases. & SIR's distribution-reweight strategy introduces higher uncertainty, making it sensitive to the intensity of paraphrasing attacks.
& $\LEFTcircle$ \\ 
    \midrule
UB~\citep{xiaoniu} is more robust than TGRL to substitution attacks. & UB and TGRL are comparably robust to synonym substitution attacks; yet, UB is more vulnerable to paraphrasing attacks. &  The distribution-reweight strategy is not superior to the distribution-shift strategy. &  $\Circle$ \\ 
\bottomrule
\end{tabular}}

\label{tbl:compare}
\end{table*}

Despite recent efforts to benchmark LLM watermarkers, existing research is limited in addressing these questions. WaterBench~\citep{waterbench} primarily focuses on watermarking effectiveness;
MarkMyWords~\citep{markmyword} mainly evaluates the robustness of a specific watermarker~\citep{john}; MarkLLM~\citep{pan2024markllm} focuses on providing a platform to compare watermark detectability, basic robustness, and text quality. \citet{zhao2024sok} provides a comprehensive overview of watermarking techniques. Moreover, these studies lack an in-depth analysis of how a watermarker's design choices impact its robustness. 
Consequently, the aforementioned questions remain largely unexplored.

To bridge this gap, this work conducts a systematic study of state-of-the-art LLM watermarkers, focusing on their attack robustness. We aim to understand how various design choices affect attack resilience and identify best practices for operating watermarkers in adversarial environments. 

We develop \system, the first open-source platform dedicated to evaluating the attack robustness of LLM watermarkers in a unified and comprehensive manner. As of 5/15/2025, \system integrates 12 state-of-the-art watermarkers, 12 representative watermark removal attacks, and 8 key metrics. Moreover, \system offers a comprehensive suite of tools for in-depth robustness assessment, including next-token distribution comparisons, attack combination analyses, and what-if scenario evaluations. Leveraging \system, we empirically evaluate the attack resilience of representative LLM watermarkers, leading to many interesting findings, which challenge the conclusions in prior work, as summarized in Table\mref{tbl:compare}. We also explore how a watermarker's design choices impact its attack robustness, unveiling critical trade-offs between different types of robustness. \textcolor{black}{In addition, we provide detailed deployment and evaluation guidelines in \mref{sec:dev_guide} and \mref{sec:eva_guide}.}

\vspace{-8pt}

%% file: taxonomy.tex
\section{LLM Watermarking}

A large language model (LLM) is typically an auto-regressive model that generates the next token $x_t$ based on previous tokens $x_{< t} \triangleq x_1, \ldots, x_{t-1}$ (including its prompt), modeled as sampling from a conditional distribution $p(x_t | x_{< t})$.

Conceptually, a watermark is a pattern embedded in a given signal (e.g., text) for identifying the signal's source. In the context of LLMs, watermarks can be used to prove that a given text is LLM-generated (or even generated by specific LLMs). An LLM watermarking method (``watermarker'') comprises three components: the LLM, watermarking procedure (generator), and detection procedure (detector), as shown in Figure\mref{fig:flow}.

Typically, the generator produces watermarked texts iteratively. At each iteration, with access to a secret key $k$, the previous tokens $x_{< t}$, and the LLM's next-token distribution $p ( x_t | x_{< t})$, the generator generates a perturbed distribution $\tilde{p} ( x_t | x_{< t})$, from which the next token $x_t$ is sampled. Meanwhile, with access to a secret key $k$, the detector determines whether $M$ generates a given text. 
In symmetric schemes, $k = k'$, whereas in asymmetric ones~\citep{public-detectable-watermark}, $k \neq k'$. This study mainly focus on the symmetric case.


The current watermarkers can be categorized based on the information carried by the watermarks (e.g., one-bit versus multi-bit) and their key design choices, including context dependency, generation strategy, and detection method. The key design factors of each category are deferred to \msec{sec:taxonomies} .
\vspace{-10pt}


%% file: platform.tex
\section{Platform}
\subsection{Threat Model}
\label{sec:threatmodel}

One critical property of LLM watermarkers is their robustness against potential attacks. We assume a threat model similar to prior work~\citep{markmyword,xuandong,umd-attack,para-attack}, as shown in Figure\mref{fig:flow}. We assume the adversary has access to sample watermarked and non-watermarked texts, but cannot reproduce the watermarking procedure or interact with the detection procedure. The adversary modifies the watermarked text $\tilde{T}$ as the altered text $T'$ such that the following objectives are met: effectiveness -- $T'$ evades the watermark detector (i.e., detected as non-watermarked) and quality -- $T'$ preserves $\tilde{T}$'s original semantics.

\subsection{Metrics}
\label{sec:metrics}
{\bf Effectiveness.} At a high level, {\system} evaluates the watermark detector's accuracy in detecting watermarked texts mainly using two metrics: true positive rate (TPR) and false positive rate (FPR). TPR measures the fraction of watermarked texts detected as watermarked, while FPR measures the fraction of non-watermarked samples wrongly detected as watermarked. Formally, let $S_+$ and $S'_+$ respectively be the sets of ground-truth and detected watermarked texts ($S_-$ and $S'_-$ correspondingly).
\begin{equation}
\text{TPR} = \frac{|S_+ \cap S'_+|}{|S_+|} \qquad \text{FPR} = \frac{|S_- \cap S'_+|}{|S_-|}
\end{equation}
In \system, we plot the receiver operating characteristic (ROC) curve that measures TPR against FPR across varying settings of detection thresholds.  In particular, the area under the ROC curve (AUC) evaluates the overall effectiveness of each watermarker. Moreover, to compare two watermarkers under specific settings, we may also measure their TPRs under a fixed FPR (e.g., 1\%).

\noindent{\bf Fidelity.} To evaluate the impact of watermarking on text quality, \system employs the following metrics to measure the difference between the original text $T$ and the watermarked text $\tilde{T}$.
We employ WER (word error rate), BLEU~\citep{papineni2002bleu}, BERTScore~\citep{zhang2020bertscore} and P-SP~\citep{wieting2021paraphrastic} as metrics to assess fidelity. Detailed descriptions of these metrics can be found in \msec{sec:fidelity_metrics}.
Note that these metrics are also used to measure the impact of watermark removal attacks on the quality of the modified text $\tilde{T}'$, relative to the watermarked text $\tilde{T}$. 

\noindent{\bf Robustness.} 
To evaluate a watermarker's attack resilience, \system measures the attack's impact on the watermarking effectiveness. Specifically, let $\tilde{T}$ and $\tilde{T}'$ respectively denote the watermarked text before and after the attack. \system compares the detector's TPR, FPR, and AUC with respect to $\tilde{T}$ and $\tilde{T}'$. Intuitively, a smaller difference indicates that the watermarker is less attack-sensitive.








%% file: eval.tex
\section{Evaluation}
\label{sec:eval}

We leverage \system to empirically assess representative LLM watermarkers, focusing on their attack robustness and other related criteria. 

%% file: attack.tex
\subsection{Experimental Setting}

\textcolor{black}{To comprehensively evaluate the robustness of LLM watermarkers, we assess all methods listed in Table~\mref{tbl:taxonomy} against the full range of attacks defined in Table~\mref{tbl:attack-taxonomy}.}

\textcolor{black}{We include three LLMs of varying scales(\mie OPT-1.3B (small), LLaMA3-7B-chat (medium), and Qwen2.5-14B-Instruct (large)) and five datasets spanning different styles and domains: C4 (text completion), HC3 (QA), Story Completion (creative writing), Law Stack Exchange (legal), and Paper Conclusion (academic summarization). This design ensures broad and unbiased robustness assessment.}

\textcolor{black}{All watermarkers and attacks are implemented following their official documentation. Table~\ref{tbl:default_parameter} summarizes the default parameter configurations used in our experiments. We further validate the fidelity and effectiveness of our implementations in Sections~\mref{sec:effectiveness} and~\mref{sec:fidelity}, confirming consistency with results reported in the original works. The evaluation results across various datasets are presented in Table~\ref{tbl:diff_dataset}. The influence of generation length is discussed in Section~\mref{sec:length}, and detailed descriptions of each attack type are provided in Appendix~\mref{appendix:watermark_removal}.}

\textcolor{black}{We report the true positive rates (TPRs) of all watermarking methods under each attack type, with the false positive rate (FPR) consistently fixed at 1\%.}

\begin{table*}[!ht]\small
\centering
\small
\def\arraystretch{1.2}
\setlength{\tabcolsep}{1.5pt}
\definecolor{myblue}{rgb}{0.8,0.8,1}
\definecolor{myred}{rgb}{1,0.8,0.8}
\caption{\small \textcolor{black}{Attack resilience of LLM watermarkers. The intensity of red shading indicates higher values, while the intensity of blue shading indicates lower values, with 0.5 serving as the threshold between the two color gradients. References: TGRL \cite{john}, UG \cite{xuandong}, UPV \cite{aiweia}, SIR \cite{aiweib}, RDF \cite{rohith}, UB \cite{xiaoniu}, DIP \cite{dipmark}, GO \cite{scott}, SynthID \cite{dathathri2024scalable}, EWD \cite{lu2024entropy} \label{tbl:attack_results_qwen}}}
\resizebox{0.98\textwidth}{!}{
\begin{tabular}{lccccccccccccccc}
    \toprule
\multirow{2}{*}{\makecell{Water\\marker}} & \multirow{2}{*}{CLEAN} & \multicolumn{3}{c}{Linguistic variation} & \multicolumn{4}{c}{Lexical editing} & \multicolumn{4}{c}{Text-mixing} & \multicolumn{3}{c}{Paraphrasing} \\

\cmidrule(lr){3-5} \cmidrule(lr){6-9} \cmidrule(lr){10-13} \cmidrule(lr){14-16}

 &  & Contra & Expan & LowCase & Swap & Typo & Syno & Missp & CP1-10 & CP3-10 & CP1-25 & CP3-25 & DP-20 & DP-40 & Trans \\ 

\midrule
TGRL & \cellcolor{myred!99}0.993 & \cellcolor{myred!94}0.944 & \cellcolor{myred!94}0.944 & \cellcolor{myred!83}0.831 & \cellcolor{myblue!14}0.429 & \cellcolor{myblue!44}0.222 & \cellcolor{myred!77}0.887 & \cellcolor{myred!82}0.915 & \cellcolor{myblue!100}0.000 & \cellcolor{myblue!67}0.667 & \cellcolor{myred!67}0.833 & \cellcolor{myred!67}0.833 & \cellcolor{myblue!67}0.667 & \cellcolor{myblue!30}0.485 & \cellcolor{myblue!44}0.222 \\ 
UG & \cellcolor{myred!99}0.993 & \cellcolor{myred!86}0.857 & \cellcolor{myred!83}0.833 & \cellcolor{myred!98}0.976 & \cellcolor{myred!93}0.929 & \cellcolor{myred!93}0.930 & \cellcolor{myred!98}0.976 & \cellcolor{myred!62}0.738 & \cellcolor{myred!86}0.857 & \cellcolor{myred!71}0.714 & \cellcolor{myred!71}0.714 & \cellcolor{myred!99}0.991 & \cellcolor{myred!98}0.976 & \cellcolor{myred!88}0.877 & \cellcolor{myred!92}0.921 \\ 
UPV & \cellcolor{myblue!20}0.400 & \cellcolor{myblue!20}0.400 & \cellcolor{myblue!20}0.400 & \cellcolor{myblue!14}0.430 & \cellcolor{myblue!84}0.080 & \cellcolor{myblue!100}0.000 & \cellcolor{myblue!24}0.380 & \cellcolor{myblue!28}0.360 & \cellcolor{myblue!100}0.000 & \cellcolor{myblue!100}0.000 & \cellcolor{myblue!100}0.000 & \cellcolor{myblue!100}0.000 & \cellcolor{myblue!60}0.200 & \cellcolor{myblue!84}0.080 & \cellcolor{myblue!52}0.240 \\ 
RDF & \cellcolor{myred!100}0.999 & \cellcolor{myred!100}0.998 & \cellcolor{myred!100}0.996 & \cellcolor{myred!100}0.996 & \cellcolor{myred!98}0.976 & \cellcolor{myred!98}0.979 & \cellcolor{myred!99}0.991 & \cellcolor{myred!99}0.993 & \cellcolor{myred!87}0.872 & \cellcolor{myred!89}0.893 & \cellcolor{myred!93}0.932 & \cellcolor{myred!98}0.978 & \cellcolor{myred!91}0.905 & \cellcolor{myred!69}0.738 & \cellcolor{myred!98}0.978 \\ 
UB & \cellcolor{myred!97}0.980 & \cellcolor{myred!100}1.000 & \cellcolor{myred!100}1.000 & \cellcolor{myred!96}0.962 & \cellcolor{myblue!100}0.000 & \cellcolor{myblue!96}0.033 & \cellcolor{myred!92}0.921 & \cellcolor{myred!100}1.000 & \cellcolor{myblue!98}0.018 & \cellcolor{myblue!95}0.042 & \cellcolor{myblue!100}0.000 & \cellcolor{myblue!51}0.485 & \cellcolor{myred!51}0.514 & \cellcolor{myblue!89}0.103 & \cellcolor{myblue!74}0.255 \\
SIR & \cellcolor{myred!98}0.978 & \cellcolor{myblue!58}0.580 & \cellcolor{myblue!50}0.500 & \cellcolor{myblue!40}0.420 & \cellcolor{myblue!36}0.320 & \cellcolor{myred!72}0.320 & \cellcolor{myblue!44}0.440 & \cellcolor{myblue!56}0.560 & \cellcolor{myblue!32}0.340 & \cellcolor{myblue!36}0.360 & \cellcolor{myblue!36}0.360 & \cellcolor{myblue!46}0.460 & \cellcolor{myblue!40}0.400 & \cellcolor{myblue!60}0.300 & \cellcolor{myblue!100}0.000 \\ 
GO & \cellcolor{myred!100}0.996 & \cellcolor{myred!100}0.996 & \cellcolor{myred!100}0.996 & \cellcolor{myred!99}0.994 & \cellcolor{myred!98}0.982 & \cellcolor{myred!96}0.956 & \cellcolor{myred!99}0.986 & \cellcolor{myred!100}0.996 & \cellcolor{myred!86}0.864 & \cellcolor{myred!89}0.887 & \cellcolor{myred!95}0.946 & \cellcolor{myred!98}0.982 & \cellcolor{myblue!60}0.640 & \cellcolor{myblue!56}0.560 & \cellcolor{myblue!67}0.667 \\ 
DIP & \cellcolor{myred!98}0.978 & \cellcolor{myred!86}0.857 & \cellcolor{myred!83}0.833 & \cellcolor{myred!87}0.867 & \cellcolor{myred!89}0.889 & \cellcolor{myred!92}0.920 & \cellcolor{myred!89}0.890 & \cellcolor{myred!88}0.879 & \cellcolor{myblue!31}0.343 & \cellcolor{myblue!4}0.478 & \cellcolor{myblue!7}0.466 & \cellcolor{myred!79}0.788 & \cellcolor{myred!69}0.685 & \cellcolor{myred!78}0.781 & \cellcolor{myred!58}0.577 \\
\tiny{TGRL+EWD} & \cellcolor{myred!100}0.995 & \cellcolor{myred!95}0.950 & \cellcolor{myred!95}0.954 & \cellcolor{myred!85}0.847 & \cellcolor{myred!65}0.650 & \cellcolor{myblue!6}0.472 & \cellcolor{myred!89}0.894 & \cellcolor{myred!93}0.931 & \cellcolor{myblue!93}0.035 & \cellcolor{myred!71}0.712 & \cellcolor{myred!84}0.843 & \cellcolor{myred!85}0.850 & \cellcolor{myred!79}0.789 & \cellcolor{myred!57}0.573 & \cellcolor{myblue!8}0.459 \\
SynthID & \cellcolor{myred!100}0.998 & \cellcolor{myred!94}0.935 & \cellcolor{myred!93}0.932 & \cellcolor{myred!86}0.856 & \cellcolor{myblue!17}0.415 & \cellcolor{myblue!36}0.321 & \cellcolor{myred!92}0.919 & \cellcolor{myred!93}0.928 & \cellcolor{myblue!92}0.039 & \cellcolor{myred!59}0.592 & \cellcolor{myred!83}0.829 & \cellcolor{myred!82}0.819 & \cellcolor{myred!65}0.654 & \cellcolor{myblue!0}0.498 & \cellcolor{myblue!54}0.232 \\
\midrule

\end{tabular}}
\end{table*}

\subsection{Robustness -- Observational Study}
Our comprehensive cross-model and cross-dataset analysis, as detailed in Table~\mref{tbl:full_attack}, demonstrates a consistent ranking of robustness across various models and datasets. This consistency indicates that the inherent algorithmic design of the watermarkers is the primary factor influencing their relative robustness, rather than model-specific or dataset-specific factors.

To provide a thorough evaluation, we selected the Qwen2.5-14B-Instruct model with the Paper Conclusion Dataset as our primary evaluation setting. As shown in Table~\mref{tbl:attack_results_qwen}, this combination effectively captures the observed robustness patterns, making it a reliable basis for our main results. This choice reflects the broader trends identified in our extensive cross-analysis efforts. We also integrated and analyzed the Entropy-based Watermark Detection (EWD)~\citep{lu2024entropy} method applied to TGRL~\citep{john}.

\subsubsection{Linguistic Variation Attack} 
\label{sec:linguistic}

This attack perturbs the linguistic features of the watermarked text, without changing its semantics.

i) Overall, most watermarkers show strong resilience against linguistic variation attacks. For instance, TGRL reaches close to 100\% TPRs under all three attacks. ii) UPV is the watermarker marginally susceptible to such attacks. This can be explained by the fact that its neural network-based detector primarily depends on implicit textual features, which appear to be sensitive to changes in linguistic characteristics. iii) The SIR shows notably inferior performance. While it employs a neural network to predict token-specific logit perturbations that are designed to be unbiased (no preference for particular tokens) and balanced (with perturbations summing to zero), its results are suboptimal, suggesting that it is fundamentally challenging to make accurate predictions under such perturbations. iv)SynthID-Text is a recent method from Google DeepMind. Its robustness profile against linguistic variations is largely comparable to that of the standard TGRL method.

\subsubsection{Lexical Editing Attack}

This attack modifies individual words while maintaining the watermarked text's semantics. 

i) TRGL, UB, SIR, and UPV tend to be more vulnerable to lexical editing attacks, compared with RDF and UG watermarkers. Intuitively, as text-dependent watermarkers (e.g., TRGL, UB, SIR, and UPV) use previous tokens as the context for the next token in both the watermark generator and detector, the lexical editing thus causes a mismatch between the generator and detector. ii) UB exhibits much higher vulnerability to such attacks, compared with the others. For example, its TPR drops near zero under the typoing and swapping attacks. This may be explained as follows. Recall that, to achieve unbiasedness, UB applies ``hard'' perturbation on the next-token distribution (e.g., by rejecting half of the vocabulary); thus the disruption to the previous tokens tends to cause a more significant mismatch between the generator and detector, compared with other watermarkers that employ ``soft'' perturbation (e.g., distribution shift and transform). iii) UPV and SIR exhibit similar vulnerabilities to the previous attack iv) In contrast, the evaluation of DIP reveals its significant advantages. It shows markedly superior resilience against lexical editing attacks, such as Swap (0.889 vs. TGRL's 0.429 TPR) and Typo (0.920 vs. TGRL's 0.222 TPR), reinforcing the value of stealthy watermark embedding. v) Similarly, our experiments with EWD method highlight the critical importance of advanced detection schemes. The key design of EWD is to make a token's influence during detection proportional to its entropy. This approach is particularly effective against attacks like Typo and Swap. Our results show that applying EWD to TGRL-watermarked text more than doubled the TPR under the Typo attack (from 0.222 to 0.472) and significantly improved resilience to Swap attacks (from 0.429 to 0.650). vi) SynthID-Text shows a marginal improvement over TGRL against Typo attacks.

\subsubsection{Text-Mixing Attack}
\label{sec:text-mixing}

This class of attacks ``dilutes'' the watermark by mixing the watermarked text with non-watermarked
text fragments. Here, to evaluate the resilience of different watermarkers against text-mixing attacks, we use the copy-pasting attack~\citep{para-attack} as the concrete attack, which embeds the watermarked text into the context of non-watermarked, human-written text (generated under the same prompt). We use CP-$n$-$m$ to denote the attack in which the modified text $T'$ consists of  $n$ segments of watermarked texts, each of length $m$\% of $|T'|$, and the rest as non-watermarked text. 

i) Overall, most watermarkers experience significant TPR drops, especially under CP-$1$-$10$ that only preserves 10\% of the watermarked text. ii) Among all the watermarkers, GO and RDF show significantly higher attack resilience. This can be attributed to their sampling strategy: both employ distribution transform, which generates the next token deterministically conditional on a given random permutation. Thus, GO and RDF tend to have stronger per-token signals than the other watermarkers that sample the next token from a given pool (e.g., green list). This observation is consistent with that in \msec{sec:length}. iii) Meanwhile, UB and UPV are the most vulnerable to the copy-pasting attack, with close to zero TPRs under CP-$1$-$10$ and CP-$1$-$25$. This can be explained as follows. The model-assisted detector of UPV determines the given text as watermarked based on its aggregated features (rather than per-token statistics), while the injected non-watermarked segments may greatly disrupt such features. Meanwhile, UB applies hard perturbation on the next-token distribution (e.g., by rejecting half of the vocabulary); thus the disruption to the previous tokens causes a significant mismatch between the generator and detector. iv) DiPmark also demonstrates much stronger performance against text-mixing attacks compared to earlier methods, maintaining high TPRs even when a large portion of the text is replaced.

\subsubsection{Paraphrasing Attack}
\label{sec:parapharse}
This class of attacks employs an additional LLM (i.e., paraphraser) to re-write the given watermarked text $\tilde{T}$ (while preserving its semantics) to evade the detector. Here, we consider the Dipper attack~\citep{retrieval-defense} as the paraphraser that rewrites $\tilde{T}$ in one shot, denoted as DP-$l$-$o$, where lexical change ($l$) indicates that $l$\% of the given text is paraphrased and order change ($o$) indicates that $o$\% of the text is re-ordered. Further, we also consider the translating attack uses a translator model Seamless-m4t-v2-large~\citep{seamlessm4t} that first translates $\tilde{T}$ to French and then translates it back. 

i) UPV and UB exhibit higher vulnerability to the Dipper attack, compared to other watermarkers. UPV's TPR drops to around zero under DP-$40$, which aligns with our analysis in \msec{sec:text-mixing}: the paraphrased text segments may greatly disrupt the aggregated textual features for UPV's model-assisted detector, while the disruption to the previous tokens may cause a substantial mismatch between UB's generator and detector, due to its rigid perturbation to the next-token distribution (e.g., rejecting half of the vocabulary). ii) In contrast, RDF and UG are especially robust against the Dipper attack. This can be attributed to their index-dependent and context-free designs, which are less sensitive to the change of previous tokens than text-dependent watermarkers (e.g., TGRL, SIR, and GO). iii) Interestingly, RDF is more vulnerable to the translating attack than the Dipper attack. This susceptibility arises from RDF's detection mechanism, which is highly sensitive to text length variations, a weakness readily exploited by the translating attack's tendency to produce shorter output text. iv) DiPmark once again shows its robustness, demonstrating strong performance against various paraphrasing attacks. In contrast, SynthID-Text's resilience in this category is similar to the TGRL baseline.

\subsubsection{Fidelity Preservation of Attacks}
\begin{figure}[!ht]
   \setlength{\abovecaptionskip}{3pt}  
    \centering
    \includegraphics[width=\linewidth]{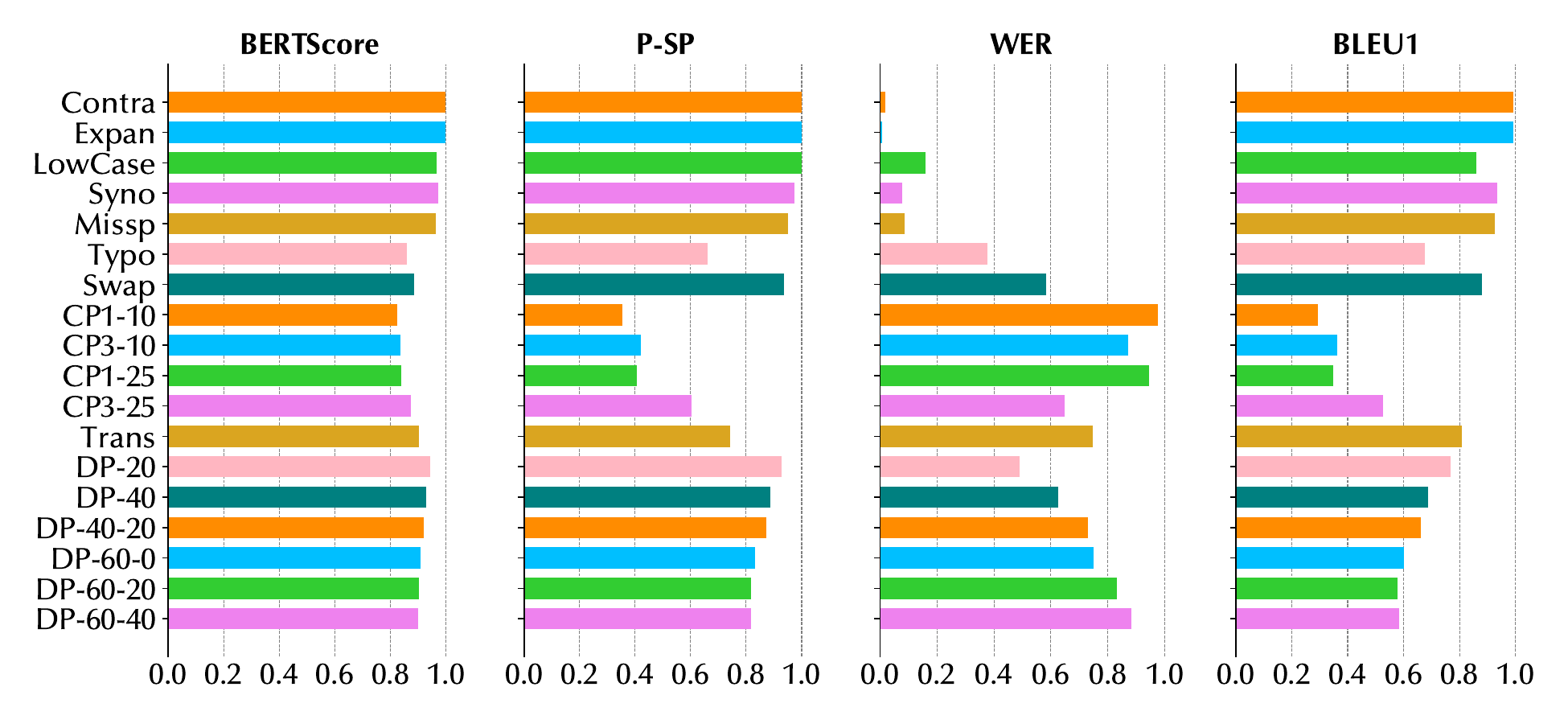}
    \caption{Quality preservation of different attacks.}
    \label{fig:exp_attack_sentiment_small}
\end{figure}

Recall that besides their attack effectiveness, another key metric for watermark removal attacks is whether they can preserve the quality of original texts. We thus compare the semantics of watermarked text $\tilde{T}$ and modified text $T'$ using the metrics in \msec{sec:metrics}. 
Figure\mref{fig:exp_attack_sentiment_small} illustrates the quality preservation of different attacks on GO, with similar results on other watermarkers, with more results in \msec{sec:appendix_sentiment}.

Observe that most attacks preserve the semantics of the watermarked text $\tilde{T}$ in the modified text $\tilde{T}'$, as measured by BERTScore and P-SP scores. In comparison, the copy-pasting (CP) attack causes more significant text-quality degradation than other attacks, in that it may disrupt the orders of watermarked and non-watermarked segments and insert duplicate segments. Also, note that most attacks emphasize the semantic similarity between $\tilde{T}$ and $\tilde{T}'$ rather than their lexical similarity (as measured by WER and BLEU scores).

\subsection{Robustness – Controlled Comparative Analysis}
\label{sec:causal}
\begin{figure}[!ht]
   \setlength{\abovecaptionskip}{5pt}  
    \centering
    \includegraphics[width=\linewidth]{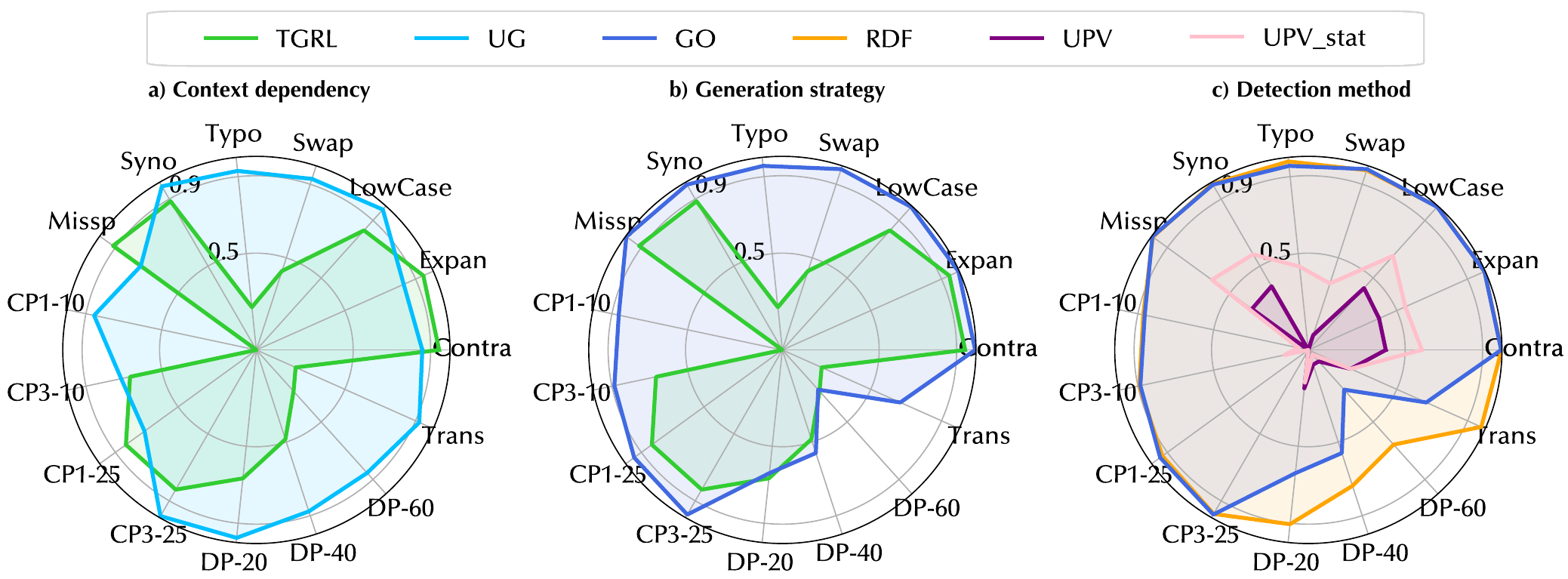}
  
    \caption{\small \textcolor{black}{Watermarker robustness to multi-attacks. a) Context dependency: TGRL (text-dependent) and UG (context-free);  b) Generation strategy: TGRL (distribution-shift) and GO (distribution-transform); c) Detection method: UPV (Model-based) and UPV$_{stat}$ (Score-based). RDF and GO.}}
    \label{fig:casual}
\end{figure}

In addition to the observational studies, we further consider conducting controlled comparative analysis to understand the impact of individual design choices (e.g., samplers). However, this is challenging in our context because the various components of a watermarker are often highly interconnected and difficult to decouple. To address this challenge, we select two watermarkers with their only difference in the design of one specific component (e.g., context dependency). The results are shown in Figure\mref{fig:casual}.

\subsubsection{Context Dependency}

\textcolor{black}{We select TGRL and UG to represent text-dependent and context-free designs, respectively. TGRL uses the previous $k$ tokens as a randomness seed to divide the vocabulary into red/green lists for the current token, whereas UG uses fixed red/green lists across all positions.}

\textcolor{black}{As shown in (a), UG consistently outperforms TGRL under most attack categories. This includes not only strong paraphrasing attacks such as DP-60 and DP-40, but also lexical editing attacks (e.g.,  Typo, Syno) and text-mixing attacks (e.g., CP1-10, CP3-10). In contrast, TGRL exhibits a noticeable drop in true positive rate (TPR) under these challenging perturbations, especially when token ordering and surface forms are significantly altered.}

\textcolor{black}{This performance discrepancy is expected. Attacks like Dipper-style paraphrasing and text mixing disrupt the continuity and order of input tokens, breaking the alignment assumed by TGRL’s context-dependent randomness during detection. In contrast, UG’s context-free design applies consistent token-level watermarking that remains robust regardless of token rearrangement or local corruption. These results further validate prior observations in UG\mcite{xuandong} regarding its robustness to syntax-level transformations.}

\subsubsection{Generation Strategy}

\textcolor{black}{Both TGRL and GO adopt context-dependent watermarking. However, they differ in their generation mechanisms: TGRL employs a distribution-shift approach with soft constraints, whereas GO utilizes a deterministic distribution-transform method on token logits.}

\textcolor{black}{As observed in the (b), GO consistently outperforms TGRL under text-mixing attacks, including CP1-10, CP3-10, and CP3-25. While TGRL’s true positive rate (TPR) drops significantly under these perturbations, GO maintains high detection performance across all copy-paste scenarios. GO also shows moderate robustness under paraphrasing attacks such as DP-40 and DP-60, where TGRL again underperforms.}

\textcolor{black}{The key difference lies in signal strength: GO's transform-based sampling embeds consistent token-level signals resilient to inserted distractors. TGRL’s soft sampling is more vulnerable to dilution and misalignment from non-watermarked fragments.}

\subsubsection{Detection Method}

\textcolor{black}{UPV supports both model-based and score-based detection. While the original work suggests model-based detection improves robustness against paraphrasing, our results challenge this claim. As shown in the radar plot (right), both UPV (model-based) and UPV$_{stat}$ (score-based) exhibit near-zero TPR under strong paraphrasing (DP-60, DP-40) and text-mixing (CP3-25), indicating a critical failure in handling high-intensity attacks. Interestingly, UPV$_{stat}$ slightly outperforms UPV under milder perturbations such as misspelling and synonym substitution. This implies that model-based detection introduces additional variance and is more prone to subtle distributional shifts, whereas score-based detection retains better stability in cleaner settings. However, neither variant withstands red-team scenarios effectively, calling into question the overall practicality of UPV’s detection mechanism.}

\textcolor{black}{Both RDF and GO adopt distribution-transform generation with score-based detection, but differ in scoring criteria. RDF leverages edit-based scoring and demonstrates consistently higher robustness than GO across most attacks. In particular, RDF shows greater resilience under strong paraphrasing (DP-40, DP-60) and text rearrangement (CP3-25), highlighting the benefits of edit-distance awareness in detecting structurally altered text.}

\subsection{Cross-analysis Between the Fidelity and Robustness}

Beyond evaluating robustness, it is crucial to consider the impact on text quality detailed in \mref{sec:fidelity}. We now give the cross-analysis between these two factors.

A fundamental trade-off between a watermark's adversarial robustness and its preservation of text fidelity is evident when analyzing the evaluated methods. Our findings suggest that the watermarkers generally fall into two distinct groups, largely defined by their underlying generation strategy. The first group, which includes TGRL, UG, SIR, and UPV, prioritizes high fidelity. These methods achieve high scores on metrics like MAUVE, indicating that the watermarked text distribution remains very close to the original. This is often due to their "distribution-shift" generation strategy, which applies a "soft" modification to the logits. However, this high fidelity comes at the cost of robustness; these methods, particularly UPV and SIR, show significant vulnerability to text-mixing and paraphrasing attacks, with their True Positive Rates (TPRs) often dropping to near zero.

In contrast, the second group, featuring RDF and GO, demonstrates a clear prioritization of robustness over fidelity. These watermarkers are exceptionally resilient to strong attacks like text-mixing and paraphrasing, maintaining high TPRs where other methods fail. This superior robustness is a direct result of their "distribution-transform" generation strategy, which embeds a stronger, more deterministic signal into the text. The consequence of this stronger signal is a noticeable decrease in fidelity, as reflected by their significantly lower MAUVE and P-SP scores. UB presents an interesting outlier; its "distribution-reweight" strategy results in low fidelity comparable to RDF and GO, but it fails to achieve similar robustness, showing extreme vulnerability to simple lexical edits. This illustrates that the nature of the design trade-off is not merely about quality degradation, but about how the specific watermarking algorithm impacts the text structure. Ultimately, the choice of a watermarker requires a deliberate decision based on the application's threat model: one must either accept lower security to preserve text quality or sacrifice fidelity for higher resilience against determined adversaries.

%% file: discussion.tex
\section{Discussion}
\label{sec:discussion}

Next, we examine the current practices of operating watermarkers in adversarial environments and explore potential improvements.

\subsection{Specific vs. Generic Detector}
\label{sec:generic-detector}

\begin{figure}[!ht]
   \setlength{\abovecaptionskip}{5pt}  
    \centering
    \includegraphics[width=1.0\linewidth]{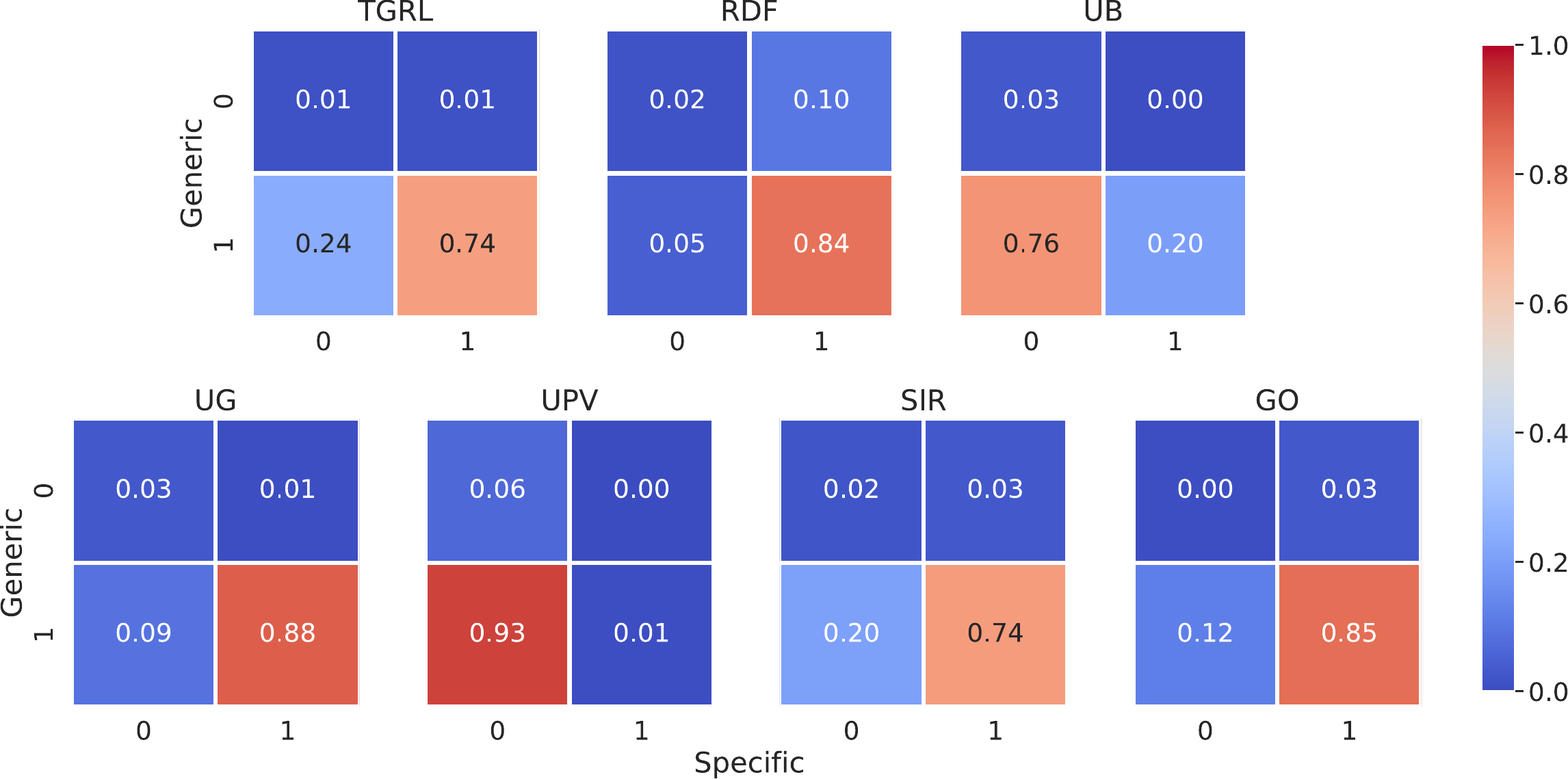}
    \caption{\small Detection of watermarked texts by watermarker-specific and generic detectors (`1' or `0' indicate that the detector detects the given watermarked text as watermarked or non-watermarked).
    \label{fig:exp_attack_sentiment-confusion}}
\end{figure}

For each watermarker, we mainly use its specific detector to detect watermarked texts. Here, we explore a generic, neural network-based detector as an alternative. To this end, 
we employ a pre-trained RoBERTa model and fine-tune it as a binary classifier using watermarked and non-watermarked texts. We use OPT-1.3B as the underlying LLM and C4 as the reference dataset. The watermarked text is generated by the LLM and the watermarker jointly, whereas the non-watermarked text is produced by ChatGPT using the same prompt to mimic human response. Table\mref{tbl:generic_eva_acc} shows the TPRs of watermarkers with generic detectors (with FPRs fixed as 1\%).

\begin{table}[!ht]\small
\centering
\renewcommand{\arraystretch}{1.25} 
\setlength{\tabcolsep}{1pt}       
\caption{\small TPRs of watermarkers with generic detectors (with FPRs fixed as 1\%).}
\begin{tabular}{c|c|c|c|c|c|c|c}
Watermarker & TGRL & RDF & UB & UG & UPV & SIR & GO \\
\hline 
TPR & 0.985 & 0.981 & 0.989 & 0.999 & 0.997 & 0.996 & 0.970 \\
\end{tabular}
\label{tbl:generic_eva_acc}
\end{table}

We compare the attack resilience of watermarker-specific and generic detectors. For each watermarker, we apply the Dipper-$40$ attack and examine whether two detectors can effectively detect watermarked texts after the paraphrasing attack. Figure\mref{fig:exp_attack_sentiment-confusion} depicts the confusion matrices of both detectors. 

We have the following findings. i) The specific and generic detectors jointly achieve a high detection rate. Across all the watermarkers, the chance that both detectors fail to detect the watermarked texts (0-0) is below 0.07. ii) For RDF, the generic detector seems less effective than the specific detector, while for UB and UPV, the generic detector outperforms the specific detector by a large margin. Recall that UB and UPV are highly sensitive to the Dipper attack (see \msec{sec:parapharse}. Thus, employing a generic detector alongside a watermarker-specific detector can be an effective strategy for enhancing the security of vulnerable watermarkers. However, note that given the availability of generic detectors, it is also feasible for the adversary to leverage such detectors as an attack checker to adapt their attacks, which we will discuss in \msec{sec:surrogate}.

\subsection{Advanced Attack}
\label{sec:adaptive}

In addition to the previous simple attacks, we investigate the effectiveness of more advanced attacks.

\subsubsection{Varying Attack Intensity}

    
\begin{table}[htbp]\small
\centering
\caption{\small Robustness of Watermarkers against Dipper Attacks with Varying Lexical and Order Changes.}
\small
\def\arraystretch{1.0}
\setlength{\tabcolsep}{5pt}
\definecolor{myblue}{rgb}{0.8,0.8,1}
\definecolor{myred}{rgb}{1,0.8,0.8}
\label{tab:dipper-robustness}
\resizebox{\linewidth}{!}{
\begin{tabular}{lcccccc}
\toprule
\multirow{3}{*}{\makecell{Water\\marker}}  & \multicolumn{6}{c}{Attack Intensity(Dipper - lexical ($l$) - order ($o$) )}\\
\cmidrule{2-7}
& 20 & 40 & 60 & 40-20 & 60-20 & 60-40 \\
\midrule
TGRL & \cellcolor{myred!95}0.952 & \cellcolor{myred!85}0.858 & \cellcolor{myred!55}0.550 & \cellcolor{myred!74}0.746 & \cellcolor{myblue!59}0.406 & \cellcolor{myred!54}0.546 \\
UG & \cellcolor{myred!94}0.940 & \cellcolor{myred!91}0.916 & \cellcolor{myred!84}0.848 & \cellcolor{myred!89}0.894 & \cellcolor{myred!84}0.846 & \cellcolor{myred!78}0.789 \\
UPV & \cellcolor{myblue!70}0.296 & \cellcolor{myblue!94}0.057 & \cellcolor{myred!64}0.645 & \cellcolor{myred!74}0.742 & \cellcolor{myred!63}0.630 & \cellcolor{myred!60}0.606 \\
RDF & \cellcolor{myred!98}0.988 & \cellcolor{myred!96}0.962 & \cellcolor{myred!84}0.840 & \cellcolor{myred!91}0.914 & \cellcolor{myred!68}0.686 & \cellcolor{myred!71}0.714 \\
UB & \cellcolor{myblue!53}0.461 & \cellcolor{myblue!85}0.149 & \cellcolor{myblue!97}0.025 & \cellcolor{myblue!93}0.070 & \cellcolor{myblue!98}0.016 & \cellcolor{myblue!99}0.009 \\
SIR & \cellcolor{myred!90}0.902 & \cellcolor{myred!81}0.814 & \cellcolor{myblue!95}0.050 & \cellcolor{myblue!93}0.065 & \cellcolor{myblue!93}0.067 & \cellcolor{myblue!94}0.051 \\
GO & \cellcolor{myred!97}0.971 & \cellcolor{myred!84}0.847 & \cellcolor{myred!50}0.500 & \cellcolor{myred!78}0.785 & \cellcolor{myblue!62}0.378 & \cellcolor{myblue!54}0.460 \\
\bottomrule
\end{tabular}
}
\end{table}

One straightforward way to improve an attack's effectiveness is to increase its intensity, potentially at the cost of other metrics (e.g., text quality). Here, we consider the Dipper attack~\citep{retrieval-defense} under varying intensity settings, denoted as DP-$l$-$o$, where lexical change ($l$) indicates that $l$\% of the given text is paraphrased and order change ($o$) indicates that $o$\% of the text is re-ordered. We compare the watermarkers' resilience under varying attack intensity, as shown in Table\mref{tab:dipper-robustness}.

i) As expected, most watermarkers observe TPR drops as the attack intensity increases. ii) Among these, UG demonstrates the most consistent resilience under varying attack intensity. This can be attributed to its context-free design: the same perturbation is applied to the next-token distribution across all the tokens, which is thus immune to the change of previous tokens. iii) Compared with text-dependent watermarkers (e.g., TGRL and GO), an index-dependent watermarker (e.g., RDF) shows stronger resilience, especially under high attack intensity (e.g., DP-$60$-$20$), due to its weaker dependency on previous tokens. iv) UPV's performance is inconsistent; it struggles with low-intensity attacks (e.g., DP-20) but shows resilience to high-strength ones (e.g., DP-60). This inconsistency can be attributed to UPV's model-assisted detector and the inherent instability of its neural network, as confirmed by repeated experiments.

\subsubsection{Combining Simple Attacks}

\begin{table}[h]\small
\centering
\small
\def\arraystretch{1.0}
\setlength{\tabcolsep}{0.9pt}
\definecolor{myblue}{rgb}{0.8,0.8,1}
\definecolor{myred}{rgb}{1,0.8,0.8}
\caption{\small \textcolor{black}{Resilience of watermarkers against individual and combined attacks(compared to single attack). \label{tbl:attack_results_opt_multi}}}
\resizebox{\linewidth}{!}{
\begin{tabular}{lccccccccc}
    \toprule

\multirow{3}{*}{\makecell{Water\\marker}} & \multicolumn{3}{c}{Contra + }&\multicolumn{3}{c}{Swap + }&\multicolumn{3}{c}{Syno + }\\
\cmidrule(lr){2-4} \cmidrule(lr){5-7} \cmidrule(lr){8-10}

 & Typo & Swap &Low & Low & Missp & Typo & Swap & Low & Typo \\ 

\midrule
TGRL & \cellcolor{myblue!41}-.046 & \cellcolor{myblue!13}-.008 & \cellcolor{myred!02}-.002 & \cellcolor{myblue!30}-.232 & \cellcolor{myblue!22}-.122 & \cellcolor{myblue!65}-.648 & \cellcolor{myblue!86}-.086 & \cellcolor{myred!03}-.020 & \cellcolor{myblue!15}-.268 \\ 
UG & \cellcolor{myblue!09}-.034 & \cellcolor{myred!04}.006 & \cellcolor{myred!08}-.009 & \cellcolor{myblue!10}-.066 & \cellcolor{myblue!01}-.002 & \cellcolor{myblue!15}-.080 & \cellcolor{myred!02}-.008 & \cellcolor{myblue!16}-.105 & \cellcolor{myblue!09}-.125 \\ 
UPV & \cellcolor{myblue!17}-.192 & \cellcolor{myblue!44}-.437 & \cellcolor{myblue!41}-.415 & \cellcolor{myblue!49}-.505 & \cellcolor{myblue!50}-.497 & \cellcolor{myblue!50}-.505 & \cellcolor{myblue!45}-.548 & \cellcolor{myblue!43}-.534 & \cellcolor{myblue!55}-.552 \\ 
RDF & \cellcolor{myblue!06}-.021 & \cellcolor{myred!03}.043 & \cellcolor{myred!03}.009 & \cellcolor{myred!02}.013 & \cellcolor{myred!01}.039 & \cellcolor{myblue!31}-.267 & \cellcolor{myred!02}-.004 & \cellcolor{myred!02}-.002 & \cellcolor{myblue!13}-.132 \\ 
UB & \cellcolor{myblue!28}-.028 & \cellcolor{myblue!16}-.016 & \cellcolor{myblue!02}-.022 & \cellcolor{myblue!02}-.028 & \cellcolor{myblue!02}-.026 & \cellcolor{myblue!03}-.088 & \cellcolor{myblue!20}-.982 & \cellcolor{myblue!02}-.258 & \cellcolor{myblue!03}-.956 \\ 
SIR & \cellcolor{myblue!49}-.524 & \cellcolor{myblue!88}-.858 & \cellcolor{myblue!86}-.840 & \cellcolor{myblue!86}-.838 & \cellcolor{myblue!86}-.866 & \cellcolor{myblue!86}-.890 & \cellcolor{myblue!86}-.862 & \cellcolor{myblue!86}-.856 & \cellcolor{myblue!86}-.894 \\ 
GO & \cellcolor{myblue!15}-.068 & \cellcolor{myblue!07}.006 & \cellcolor{myred!02}.000 & \cellcolor{myblue!18}-.258 & \cellcolor{myblue!02}-.092 & \cellcolor{myblue!86}-.834 & \cellcolor{myblue!98}-.098 & \cellcolor{myred!02}-.002 & \cellcolor{myblue!15}-.118 \\ 
\bottomrule
\end{tabular}
}
\end{table}

We first explore whether combining two attacks improves the attack's effectiveness. Here, considering the most feasible combinations, we only combine two simple attacks from the ``weak'' linguistic variation and lexical editing attacks. Table\mref{tbl:attack_results_opt_multi} illustrates the effectiveness of such combined attacks.

We have a set of interesting observations. i) UPV and SIR, which demonstrate resilience against all simple attacks, are highly vulnerable to all the combined attacks. For instance, the TPR of SIR drastically drops to below 0.3 under the contracting+typoing attack. ii) UB, which is vulnerable to the typoing and swapping attacks, is consequently vulnerable to all the combined attacks that involve typoing or swapping. iii) TGRL and GO, which are robust against all the simple attacks (including typoing and swapping), show significant vulnerability to the typoing+swapping attack. This can be explained by that as typoing and swapping respectively disrupt the tokenization and token-indexing, their combination may substantially amplify such effects. iv) RDF and UG are especially robust against the combined attacks. This can be attributed to their ``weaker'' context dependencies, which is consistent with the findings in \msec{sec:parapharse}.







\vspace{-10pt}
\subsubsection{Adaptive Attack} 
\label{sec:surrogate}

\begin{figure}[!ht]
    \centering
    \includegraphics[width=0.6\linewidth]{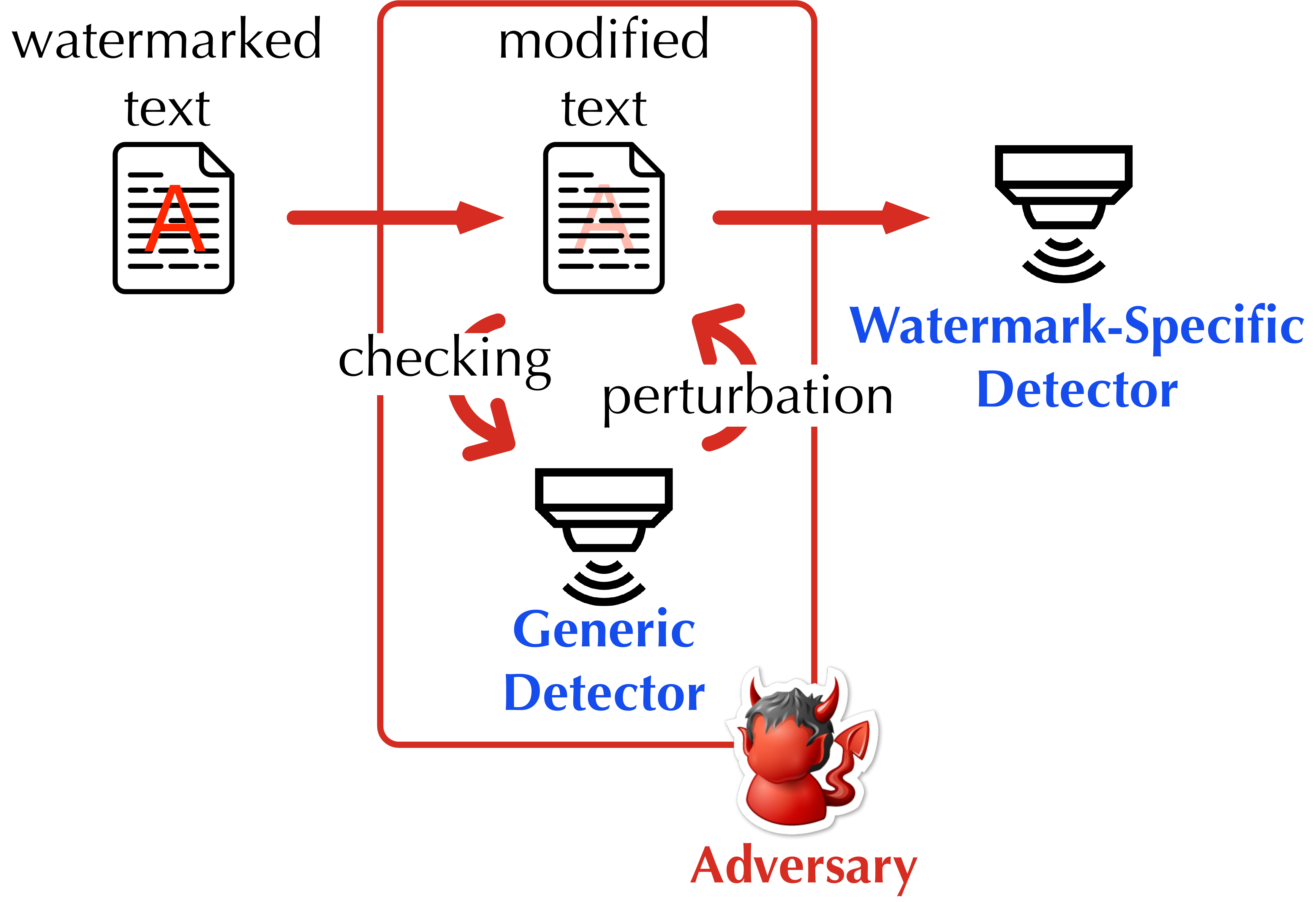}
    \caption{\small Attacks leveraging surrogate detectors. \label{fig:generic-detector}}
\vspace{-4pt}
    
\end{figure}


Given the availability of generic detectors, it is possible for the adversary to exploit such detectors to adapt their attacks. We consider a scenario as shown in Figure\mref{fig:generic-detector}: the adversary performs a gradient-based  attack~\citep{gradient_based_attack} that iteratively modifies the watermarked text to evade the generic detector, and then forwards the modified text to the target, watermark-specific detector.



\begin{figure}[!ht]
   \setlength{\abovecaptionskip}{5pt}  
    \centering
    \includegraphics[width=\linewidth]{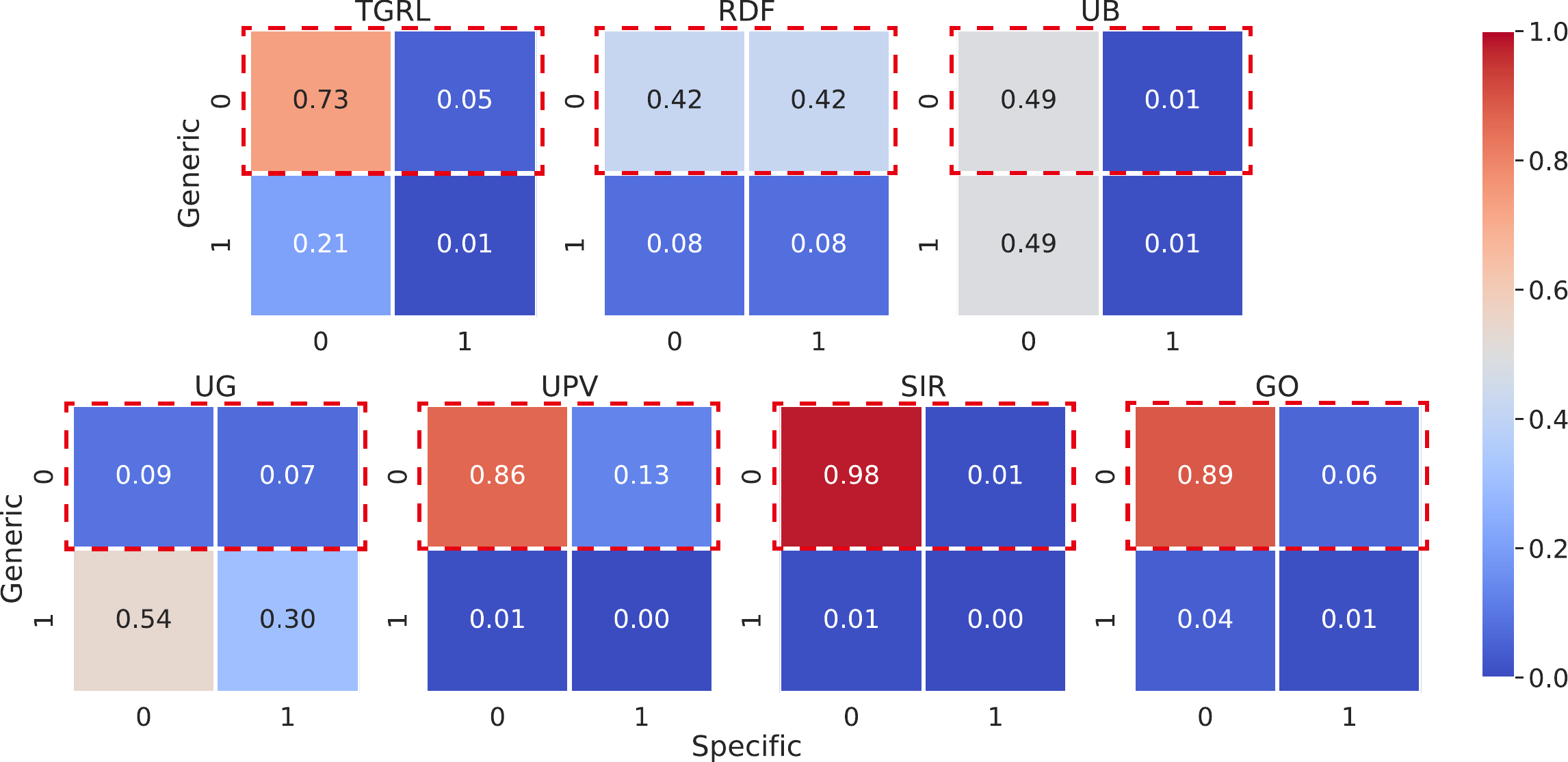}
    \caption{\small Watermark detection by watermarker-specific and generic detectors on gradient-based attacked samples (`1' or `0' indicate that the detector detects the given watermarked text as watermarked or non-watermarked). \label{fig:adver-attack-res}}
\vspace{-4pt}
\end{figure}


We evaluate this attack's effectiveness using 500 watermarked texts from the C4 dataset. We use GBDA~\citep{gradient_based_attack} as the adversarial attack and limit the steps of perturbations to 100. Compared to Dipper, GBDA makes more substantial textual modifications. The results are summarized in Figure\mref{fig:adver-attack-res}. i) Leveraging the surrogate detector significantly improves the attack effectiveness: with the BERTScore and BLEU-1 between $\tilde{T}$ and $\tilde{T}'$ is about 0.764 and 0.188, respectively, slightly lower than the Dipper attack, the detection rates of most watermarkers drop below 10\%.  ii) Although some samples do not evade the generic detector, they evade the specific detector successfully (e.g., TGRL, UG, and UB). iii) UG and RDF exhibit greater robustness than the other watermarkers. Specifically, for RDF, 42\% of the samples evade the generic detector but are still detected by the RDF-specific detector. This superior robustness is likely due to their weaker context dependencies.

\subsubsection{Leveraging Expert LLMs}

\begin{figure}[h]
   \setlength{\abovecaptionskip}{5pt}  
    \centering
    \includegraphics[width=0.9\linewidth]{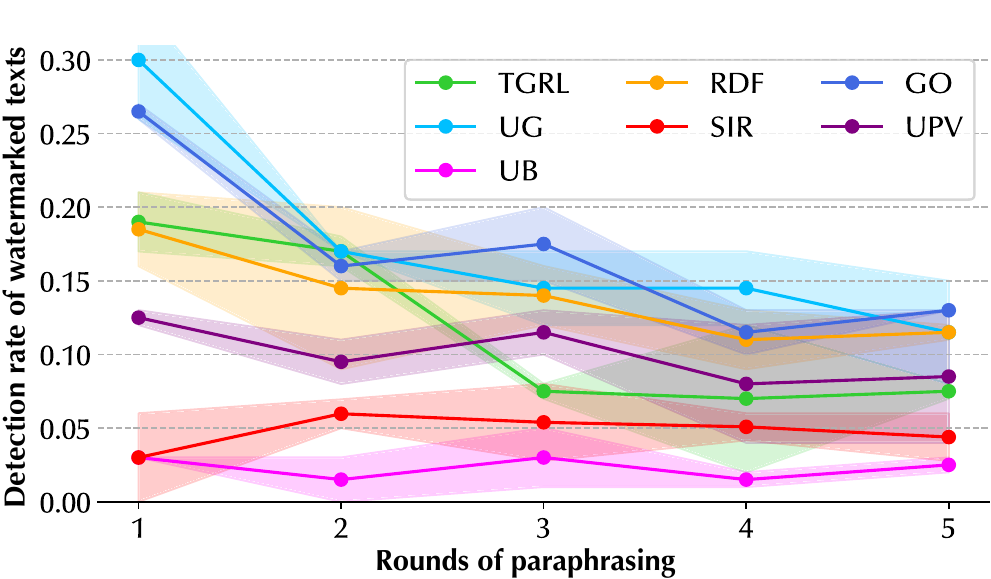}
     \caption{\small Effectiveness of paraphrasing attacks with ChatGPT as the paraphraser.}
    \label{fig:exp_attack_gpt}
\vspace{-4pt}
    
\end{figure}

We investigate the impact of highly capable LLMs in paraphrasing attacks by using ChatGPT to rewrite watermarked texts. For each watermarker, we randomly sample 100 successfully detected watermarked texts and query ChatGPT with  prompt: “Paraphrase the following text and keep length similar”.  To assess the effect of paraphrasing strength, we also apply up to 5 rounds of iterative paraphrasing, following the strategy in~\citep{watermark-sand}, which originally tested up to 300 rounds using T5. We focus on practical, resource-limited attackers, while their work~\citep{watermark-sand} explores worst-case robustness. Results (Figure~\mref{fig:exp_attack_gpt}) show that:
i) A single round of ChatGPT paraphrasing reduces detection rates of all watermarkers below 0.3, demonstrating that expert LLMs pose a strong threat. However, as noted in \msec{sec:threatmodel}, such powerful LLMs may fall outside our target threat model.
ii) For more resilient methods (e.g., UG and GO), detection rates still drop below 0.15 after a few paraphrasing rounds, confirming similar trends observed in~\citep{watermark-sand}.

\ifthenelse{\boolean{textswitch}}
{
\subsection{Evaluation Guidelines}

Next, we propose a set of guidelines for evaluating the robustness of LLM watermarkers. These guidelines incorporate our findings in \msec{sec:eval} and \msec{sec:discussion}, providing a minimal checklist to claim the robustness of an LLM watermarker.

{\bf LLMs and tasks.} Our experiments show that the referenced watermarkers show varying robustness across different LLMs and datasets. We speculate that there exists an intricate interplay between the watermarking mechanism, the LLM's capability, and the task's complexity. We thus recommend experimenting on i) LLMs with varying capability (e.g., as measured by perplexity), ii) datasets for different tasks (e.g., summarization and question-answering), and iii) their combinations.

{\bf Attacks.} Notably, using highly capable LLMs or applying computationally expensive rewriting can easily generate highly-quality, non-watermarked texts to evade detection; however, such attacks negate the need for watermark removal attacks in the first place. We thus recommend focusing on computationally efficient attacks such as linguistic variation, lexical editing, and lightweight paraphrasing, as well as their combinations, which reflects the risks of watermark removal attacks in practical settings.

{\bf Robustness.} It is often critical to properly set the decision threshold for a watermarker (and also the attacks) to fully assess its robustness~\citep{sok-image}, which unfortunately is often missing in the original papers. To overcome this issue, we recommend i) measuring the overall effectiveness (TPR) in terms of ROC (across different threshold settings), ii) measuring the TPR under a fixed FPR (e.g., 0.01), and iii) considering varying attack intensity.

{\bf Fidelity.} It is notoriously challenging to meaningfully measure the quality of text data~\citep{pillutla2023mauve}. We recommend employing a variety of metrics (e.g., BERTScore, P-SP, MAUVE) to comprehensively measure the quality retention of watermarkers as well as attacks. In addition, one may also leverage external tools (e.g., more advanced LLMs such as GPT-4) or human evaluation to provide a more accurate assessment if feasible.

}

%% file: conclusion.tex
\vspace{-8pt}
\section{Conclusion}

In this paper, we systematize the existing LLM watermarkers and watermark removal attacks, mapping out the design space for various watermarking and attacking techniques. We then design and implement \system, the first open-source platform devoted to assessing the attack robustness of LLM watermarkers in a unified and comprehensive manner. Leveraging \system, we conduct a systematic evaluation of the robustness of existing watermarkers, addressing unresolved questions, revealing design trade-offs, and identifying opportunities for further improvement. Our findings shed light on the current LLM watermarking techniques, while \system serves as a valuable benchmark aiding future research.

\newpage

\section*{Limitations}

We now examine the limitations of our study and identify promising directions for future research.

{\bf Watermarkers.} This study primarily focuses on training-free, pre-generation watermarking, which is applicable to any given LLM and offers flexible control over multiple criteria (e.g., quality, effectiveness, and robustness). While we have included peer-reviewed watermarking methods to the best of our ability, many methods, such as training-time watermarking and post-generation watermarking, are not covered in this paper.


{\bf Threat Model.} Notably, our evaluation is based on the following assumptions about adversary capabilities: i) the adversary can modify the watermarked text in a computationally efficient manner (e.g., synonym substitution), ii) the adversary uses LLMs less capable than the target LLM, and/or iii) the adversary can train a detector using given watermarked and non-watermarked texts. We argue that these assumptions are realistic and practical. Otherwise, using more capable LLMs could easily generate high-quality, non-watermarked texts to evade detection, while launching computationally expensive attacks would significantly increase the adversary's cost. Future research directions include exploring alternative threat models to expand the scope of our analysis.

{\bf Causal Analysis of Design Choices.} While our watermarker taxonomy provides a clear classification framework, we focus on analyzing the holistic design choices of each watermarked (rather than individual design modules) to elucidate the underlying factors driving our experimental observations. This limitation stems from the relatively small number of watermarkers within each taxonomic category, precluding definitive conclusions about the vulnerability of specific taxonomic elements based on current experimental evidence.

A more granular assessment of specific design choices would ideally involve comprehensive ablation studies, systematically modifying individual design elements and comparing their performance against baseline configurations. However, this approach faces significant practical challenges due to the intricate inter-dependencies and tight coupling among watermarker components. To partially address this challenge, in \msec{sec:causal}, we strategically compare two watermarkers that differ only in the design of one specific component (e.g., context dependency), enabling the examination of the effects of specific design variations. We consider a more systematic causal analysis as future research.


{\bf Parameter Tuning. }Our comparative analysis employs AUC curves based on default parameter settings across watermarkers. For benchmarking purposes, we evaluate True Positive Rate (TPR) at a fixed False Positive Rate (FPR) of 1\%, aligning with established practices in comparative studies. This standardized approach enables systematic assessment of watermark detection effectiveness while maintaining consistent false positive control. While this approach provides a pragmatic framework for comparative evaluation, it underscores the importance of future research into comprehensive parameter optimization. Such studies could reveal the full performance potential of each method and yield deeper insights into their relative strengths and limitations. 

\section*{Ethics Considerations}

In this paper, we conduct a systematic study of state-of-the-art LLM watermarkers, focusing on their robustness against watermark removal attacks. We aim to understand how various design choices affect watermarkers' resilience and identify best practices for operating watermarkers in adversarial environments.

{\bf Stakeholder Considerations.} The primary stakeholders affected by this research include users and developers of LLM watermarkers, as well as the broader community relying on these techniques. By identifying the strengths/limitations of various watermarkers, our work could influence the perceived reliability and deployment of LLM watermarkers in critical applications. Conversely, exposing these vulnerabilities allows for the improvement of the current techniques, ultimately contributing to more secure and robust systems.

{\bf Potential Harms and Benefits.}
Exposing vulnerabilities in widely adopted security mechanisms can yield contrasting outcomes. Initially, it may erode trust in LLM watermarkers as reliable safeguards, potentially leading to their underuse and leaving systems exposed to unchecked risks. However, by illuminating these weaknesses, our research catalyzes the development of more robust techniques and fosters a nuanced comprehension of their constraints, ultimately fortifying the long-term security ecosystem.

{\bf Future Research and Mitigation.}
Our study has identified several potential countermeasures to address the vulnerabilities we uncovered. These recommendations are designed to steer future research and assist developers in bolstering the security of LLM watermarkers.

\section*{Acknowledgements}
This work is supported by the National Science Foundation under Grant No. 2405136 and 2406572, and OpenAI's Researcher Access Program. 

%% file: appendix.tex
\newpage
\appendix
\label{sec:appen}


\section{Parameter Setting}
\label{sec:parameter}
Table\mref{tbl:default_parameter} lists the default setting of the parameters of each watermarker in our evaluation. Note that $\gamma$ and $\delta$ are the gamma and delta used in the watermarker, and $n$ denotes the window size.


\begin{table}[!ht]
\small
\centering
\resizebox{0.98\linewidth}{!}{

\begin{tabular}{c|c|c}
\textbf{Watermarker} & \textbf{Parameter} & \textbf{Setting} \\
\hline
\multirow{2}{*}{TGRL} & $\gamma$ & 0.25 \\
                       & $\delta$ & 2.0 \\
\hline
\multirow{2}{*}{UG}   & $\gamma$ & 0.5 \\
                       & $\delta$ & 2.0 \\
\hline
\multirow{5}{*}{CTWL} & $\delta$ & 1.5 \\
                       & $n$ & 10 \\
                       & message code length & 20 \\
                       & encode ratio & 10.0 \\
                       & message strategy & vanilla \\
\hline
\multirow{5}{*}{UPV}  & $\gamma$ & 0.5 \\
                       & $\delta$ & 2.0 \\
                       & $n$ & 3 \\
                       & bit number & 16 \\
                       & layers & 9 \\
\hline
\multirow{1}{*}{UB}   
                       & watermark type & delta \\
\hline
\multirow{3}{*}{SIR}  & $\delta$ & 1.0 \\
                       & $n$ & 10 \\
                       & watermark type & context \\
\hline
\multirow{1}{*}{GO}   & $n$ & 3 \\
\hline
\multirow{1}{*}{RDF}  & number of random sequences & 50 \\
\end{tabular}}
\caption{Default parameter setting of watermarkers.}
\label{tbl:default_parameter}
\end{table}

\section{Taxonomies}
\label{sec:taxonomies}
We present a taxonomy of LLM watermarkers, as summarized in Table\mref{tbl:taxonomy}. The current watermarkers can be categorized based on the information carried by the watermarks (e.g., one-bit versus multi-bit) and their key design choices, including context dependency, generation strategy, and detection method. Next, we mainly focus on the key design factors.

\begin{table*}[!ht]\small
\centering
\renewcommand{\arraystretch}{1.15}
\setlength{\tabcolsep}{1.5pt}
\resizebox{0.98\textwidth}{!}{

\begin{tabular}{r|c|c|c|c|c|c|c|c|c|c}
\multirow{2}{*}{\bf \makecell{Water\\marker}} & \multirow{2}{*}{\bf \makecell{Infor\\mation}}  & \multicolumn{3}{c|}{\bf Context Dependency} &  \multicolumn{3}{c|}{\bf Generation Strategy} & \multicolumn{3}{c}{\bf Detection Method}\\
\cline{3-11}
& & Index-dep. & Text-dep. & Context-free & Dist.-shift & Dist.-reweight & Dist.-transform & Score-  & Diff.- & Model-based \\ 
\hline
TGRL & \multirow{7}{*}{one-bit} & & \faCheck & & \faCheck  & & & 
\faCheck &  &  \\
UG  & &  & & \faCheck  & \faCheck & &  & \faCheck &  &  \\
UPV & & & \faCheck &  & \faCheck  & & &  &  & \faCheck \\
SIR & & & \faCheck &   & & \faCheck & & \faCheck &  & \\
RDF& & \faCheck & & & & & \faCheck & \faCheck &  & \\
UB  & &  & \faCheck  && & \faCheck & &  & \faCheck & \\
DIP  & &  & \faCheck  && & \faCheck & &  \faCheck&  & \\
GO & & &\faCheck & & & & \faCheck  & \faCheck &  & \\
\hline
CTWL & \multirow{2}{*}{multi-bit} &  & \faCheck& & &\faCheck& &  &  & \faCheck
\\
MPAC & &  & \faCheck& & & &  \faCheck & \faCheck &  & 
\end{tabular}
}
\caption{A taxonomy of LLM watermarkers. References: TGRL \cite{john}, UG \cite{xuandong}, UPV \cite{aiweia}, SIR \cite{aiweib}, RDF \cite{rohith}, UB \cite{xiaoniu}, DIP \cite{dipmark}, GO \cite{scott}, CTWL \cite{lean}, MPAC \cite{kiyoon}.\label{tbl:taxonomy}}
\end{table*}

\subsection{Context Dependency}

The watermarker applies a perturbation $\Delta_t$ to the LLM's next-token distribution $p ( x_t | x_{ < t})$ as $\tilde{p}( x_t | x_{ < t}) =   p ( x_t | x_{ < t}) + \Delta_t$, from which the next token $x_t$ is generated. The perturbation $\Delta_t$ often depends on the given context. Three types of context dependencies are typically used in the existing watermarkers.

{\bf Index-dependent watermark.} The watermarker produces a pseudo-random number $r_t$ by applying a keyed hash function 
\begin{equation}
\label{rdf_eq}
r_t = f_k(t),
\end{equation}
that only depends on the index $t$ of the next token; $r_t$ is then used to generate the perturbation $\Delta_t$~\citep{rohith}.

{\bf Text-dependent watermark.} The watermarker considers the previous tokens $x_{<t}$ as the context window to generate $\Delta_t$~\citep{john,scott,xiaoniu}. For instance, GO~\citep{scott} 
computes the hash of the concatenation of previous $w$ tokens as $r_t$:
\begin{equation}
\label{go_eq}
r_t = f_k(x_{t-w} \| \ldots \| x_{t-1} ),
\end{equation}

while TGRL~\citep{john} suggests using the min hash of the previous token: 
\begin{equation}
r_t =  \min(f_k(x_{t-w}), \ldots, f_k(x_{t-1} ))
\end{equation}
Similarly, UB~\citep{xiaoniu} concatenates the previous $w$ tokens to generate the context code for reweighting the logits. UPV~\citep{aiweia} and SIR~\citep{aiweib} use neural networks to generate $\Delta_t$ based on the previous tokens. 

{\bf Context-free watermark.} The watermarker applies a universal perturbation $\Delta_t$ across the next-token distributions of all the tokens without considering their contexts~\citep{xuandong}.

\subsection{Generation Strategy} 

By applying the perturbation $\Delta_t$ to the LLM's next-token distribution $p ( x_t | x_{ < t})$, the watermarker generates the new distribution $\tilde{p}( x_t | x_{ < t})$ to sample the next token $x_t$. The existing sampling strategies can be categorized as follows.

{\bf Distribution shift.} 
TGRL~\citep{john} modifies $p ( x_t | x_{ < t})$ by adding a shift $\delta$ to the logits of ``green-list'' tokens (the remaining as ``red-list'' tokens) as the modified distribution $\tilde{p}( x_t | x_{ < t})$. A token $x$ is considered as green listed if $\pi_{r_t}(x) < \gamma d$ where $\pi_{r_t}$ is a permutation seeded by $r_t$, $\gamma$ is a parameter to control the size of the green list, and $d$ is the number of vocabulary size. Similarly, UG~\citep{xuandong} uses a fixed red-green split over the vocabulary, showing greater robustness than TGRL against edit-distance-bounded attacks due to its ``hard'' split. UPV~\citep{aiweia} selects the top-$k$ tokens from $p ( x_t | x_{ < t})$, applies a neural network to predict the green-list tokens from these tokens, and adds $\delta$ to the logits of green-list tokens. Unlike the other strategies, the distribution-shift strategy preserves the diversity of generated tokens; however, it can not be made indistinguishable, as $\tilde{p}( x_t | x_{ < t})$ and $p( x_t | x_{ < t})$ are inherently distinguishable.

{\bf Distribution reweight.} 
Similar to distribution shift, this strategy alters the next-token distribution but uniquely perturbs the logit of each token. For instance, SIR~\citep{aiweib} trains a neural network to predict the perturbation to logit of each token, which is unbiased (no preference over specific tokens) and balanced (with total perturbation summing up to 0). UB~\citep{xiaoniu} advocates unbiased watermarking such that the expectation of the reweighted distribution agrees with the original distribution. It proposes two reweighting schemes: $\delta$-reweighting uniformly samples a token from $p( x_t | x_{ < t})$ and changes its probability to 1; $\gamma$-reweighting shuffles all the tokens, rejects the first half, and double the probabilities of the remaining half. DIP~\citep{dipmark} also uses a distribution-reweight generation strategy similar to UB but does not need to access the LM during detection.

{\bf Distribution transform.} Another line of watermarkers apply randomized transform on $p( x_t | x_{ < t})$ to sample $x_t$. For instance, RDF-EXP~\citep{rohith} and GO~\citep{scott} use the Gumbel-max trick and apply the exponential transform. Let $p(x_t | x_{ < t}) = \{ p_i\}_{i=1}^d$ be the
distribution over the next token $x_t$. Then $x_t$ is sampled as:
\begin{equation}
x_t = \arg\max_{  i \leq d}  r_i^{1/{p_i}}
\end{equation}
where $r_i$ is generated by the pseudo-random functions in \meq{rdf_eq} and \meq{go_eq}.
Similarly, RDF-IST~\citep{rohith} applies inverse transform over $p(x_t | x_{ < t})$. With $\pi_k$ as a random permutation seeded by the secret key $k$, the next token $x_t$ is selected as:
\begin{equation}
x_t = \pi_k \left( \min_{  j \leq d}  \sum_{i=1}^j  p_{\pi_k(i)} \geq r_t \right)
\end{equation}
which is the smallest index in the inverse permutation such that the CDF of the next token distribution exceeds $r_t$. The distribution-transform strategy does not alter the next-token distribution, thus preserving the original text distribution (i.e., indistinguishability).

\subsection{Detection Method}

The watermarker's detector determines whether a given text $T = (x_1, \ldots, x_n)$ is watermarked or not. Next, we categorize the existing detection methods as follows.

{\bf Score-based detection.} The detector computes the random value $r_i$ for each position, the per-token statistics $s(x_i, r_i)$, and a score over such statistics, which is then subjected to a one-tailed statistical test to determine whether the text is watermarked. The per-token statistics vary with the concrete generators. For instance, GO~\citep{scott} defines $s(x_i, r_i) = -\log (1 -h_{r_i}(x_i))$, while TGRL~\citep{john} defines 
$s(x_i, r_i) = 1$ if $x_i$ is in the green list and 0 otherwise. One simple way to aggregate the per-token statistics is to compute their sum~\citep{scott,john,xuandong,aiweib}:
\begin{equation}
S = \sum_{i=1}^n  s(x_i, r_i)
\end{equation}
However, as the random values may be misaligned with the tokens (e.g., due to editing), a more robust way is to compute the alignment score (e.g., edit score~\citep{rohith,lean}):
\begin{align}
S &= s^\psi(n, n) \notag \\
\text{s.t.} \; s^\psi(i, j) &= \min \left\{
\begin{array}{l}
s^\psi(i-1, j -1) + s(x_i, r_j), \\
s^\psi(i, j -1) + \psi, \\
s^\psi(i-1, j) + \psi
\end{array}
\right.
\end{align}
where $\psi$ is the ``edit-cost'' parameter.

{\bf Differential-based detection.}  This line of detectors also relies on the score of a given text. However, the score is computed by comparing the given text with the non-watermarked text generated by the same LLM. For instance, UB~\citep{xiaoniu} computes the log-likelihood ratio (LLR) score:
\begin{equation}
s(i) = \log \frac{\tilde{p}(x_i|x_{< i})}{p(x_i|x_{< i})}
\end{equation}
and its more robust maximin variant. However, note that these detectors naturally require accessing the original LLM, which is not always feasible.

{\bf Model-assisted detection.} Instead of computing the per-token statistics, which are often subject to watermark removal attacks, one may also train a model to predict whether the given text is watermarked. UPV~\citep{aiweia} trains a neural network, which shares the same embedding layers with its generator, to detect watermarked texts. Similarly, one may develop a generic detector by training it to distinguish watermarked and non-watermarked texts. We explore this option in \msec{sec:eval}.

\section{Fidelity Metrics}
\label{sec:fidelity_metrics}

In this paper, to evaluate the impact of watermarking on text quality, we employs the following metrics to measure the difference between the original text $T$ and the watermarked text $\tilde{T}$. Below, we provide a detailed description of each metric.

WER (word error rate) measures the percentage of mismatched tokens between $\tilde{T}$ and $T$ relative to the total number of tokens in $T$. This is a lexical metric used to measure the fraction of tokens that have been modified.

BLEU~\citep{papineni2002bleu} measures the lexical similarity of $T$ and $\tilde{T}$ by calculating the proportion of $n$-grams matched in $\tilde{T}$ and $T$.

BERTScore~\citep{zhang2020bertscore} measures the token-level similarity of $T$ and $\tilde{T}$ by leveraging the pre-trained contextual embeddings from BERT~\citep{bert} and matching tokens in $T$ and $\tilde{T}$ using cosine similarity.


P-SP~\citep{wieting2021paraphrastic} evaluates the semantic similarity of $T$ and $\tilde{T}$ using the cosine similarity of their encodings. In particular, the encoding of $T$ (or $\tilde{T}$) is calculated by averaging the embeddings of its subword units generated by SentencePiece~\citep{sentencepiece}.


MAUVE~\citep{pillutla2023mauve} compares the distributions of $T$ and $\tilde{T}$ by computing an information divergence curve within a quantized embedding space. The area under this divergence curve provides a scalar summary of the trade-off between Type I ($T$ places high mass in areas where $\tilde{T}$ has low mass) and Type II (vice versa) errors.

Note that these metrics are also used to measure the impact of watermark removal attacks on the quality of the modified text $\tilde{T}'$, relative to the watermarked text $\tilde{T}$.

\section{Details of Watermark Removal Attacks}
\label{appendix:watermark_removal}

\begin{table}[ht]\small
\centering
\renewcommand{\arraystretch}{1.2}
\setlength{\tabcolsep}{1pt}
\resizebox{0.98\linewidth}{!}{

\begin{tabular}{rcc}
\toprule
{\bf Attack} & {\bf Category} & {\bf Resource} \\
\midrule
Lowercasing  & \multirow{3}{*}{Linguistic variation}  &  / \\
Contracting  & & / \\
Expanding  & & / \\
\midrule 
Misspelling & \multirow{4}{*}{Lexical editing}  &  Common misspellings \\ 
Typoing & & / \\
Synonymizing & & WordNet\\
Swapping & & / \\
\midrule
Copy-pasting &  Text-mixing & Non-watermarked text\\
\midrule
Deep-paraphrasing & \multirow{2}{*}{Paraphrasing} & LLM-based paraphraser  \\
Translating & & LLM-based translator \\
\midrule
\multicolumn{2}{c}{Black-box adversarial attack} & Generic detector\\
\bottomrule
\end{tabular}}
\caption{A taxonomy of watermark removal attacks.\label{tbl:attack-taxonomy}}
\end{table}

We also present a taxonomy of existing watermark removal attacks according to their underlying perturbation and required resources, as summarized in Table\mref{tbl:attack-taxonomy}.

{\bf Linguistic variation attack.} 
This class of attacks perturb the linguistic features of the watermarked text, without changing its semantics. We consider the set of perturbations in HELM~\citep{liang2022holistic}, which simulate natural linguistic variations encountered in human interactions with text typing interfaces: i) {\em Lowercasing} converts all the words to lower-cases, which potentially affects the interpretation of proper nouns or emphases. ii) {\em Contracting/expanding}  replaces phrases with their contracted/expanded forms (e.g., ``I am'' to ``I'm'' and vice versa), which may impact the tokenizer. 

{\bf Lexical editing attack.} 
This class of attacks modifies individual words, aiming to maintain the original text's semantics. Specifically, we consider the following editing operations: i) {\em Misspelling}, similar to text-bugger~\citep{text-bugger}, replaces words with their common misspellings (plural forms also considered); ii) {\em Typoing} replaces certain letters in a word with others; iii) {\em Synonymizing} replaces words with their synonyms using WordNet~\citep{miller-1994-wordnet}; and iv) {\em Swapping} randomly exchanges the positions of two words within the text, which alters the text structure while potentially preserving the overall semantics.

{\bf Text-mixing attack.}
This class of attacks aims to ``dilute'' the watermark by mixing the watermarked text with non-watermarked text fragments. Specifically, the {\em copy-pasting} attack~\citep{para-attack} embeds the watermarked text into the context of non-watermarked, human-written text. Note that the influence of non-watermarked text can be controlled by setting the fractions of watermarked and non-watermarked text fragments (i.e., the mixing weights).

{\bf Paraphrasing attack.} This class of attacks relies on an additional LLM (i.e., paraphraser) to re-write the given watermarked text to evade the detector. For instance, a light paraphraser~\citep{umd-attack} (e.g., T5-based paraphraser~\citep{t5}) can paraphrase the watermarked text sentence-by-sentence, while a more capable paraphraser (e.g., DIPPER~\citep{retrieval-defense}) paraphrases the watermarked text in one-shot, also enabling to control lexical diversity and token order diversity.

Similar to paraphrasing, the {\em translating} attack uses a translator LLM (e.g., Seamless ~\citep{seamlessm4t}) to cycle the watermarked text through multiple languages (e.g., from English to French and back to English). This process can significantly alter the sentence structure and phrasing. 

In this class of attacks, we assume the adversary has access to a generic detector 
that is trained to distinguish watermarked and non-watermarked texts. The adversary then perturbs the watermarked text based on this surrogate detector. We explore this attack type in \msec{sec:discussion}.

\begin{figure*}[ht!]
    \centering
    \setlength{\abovecaptionskip}{5pt}  
    \includegraphics[width=1.0\linewidth]{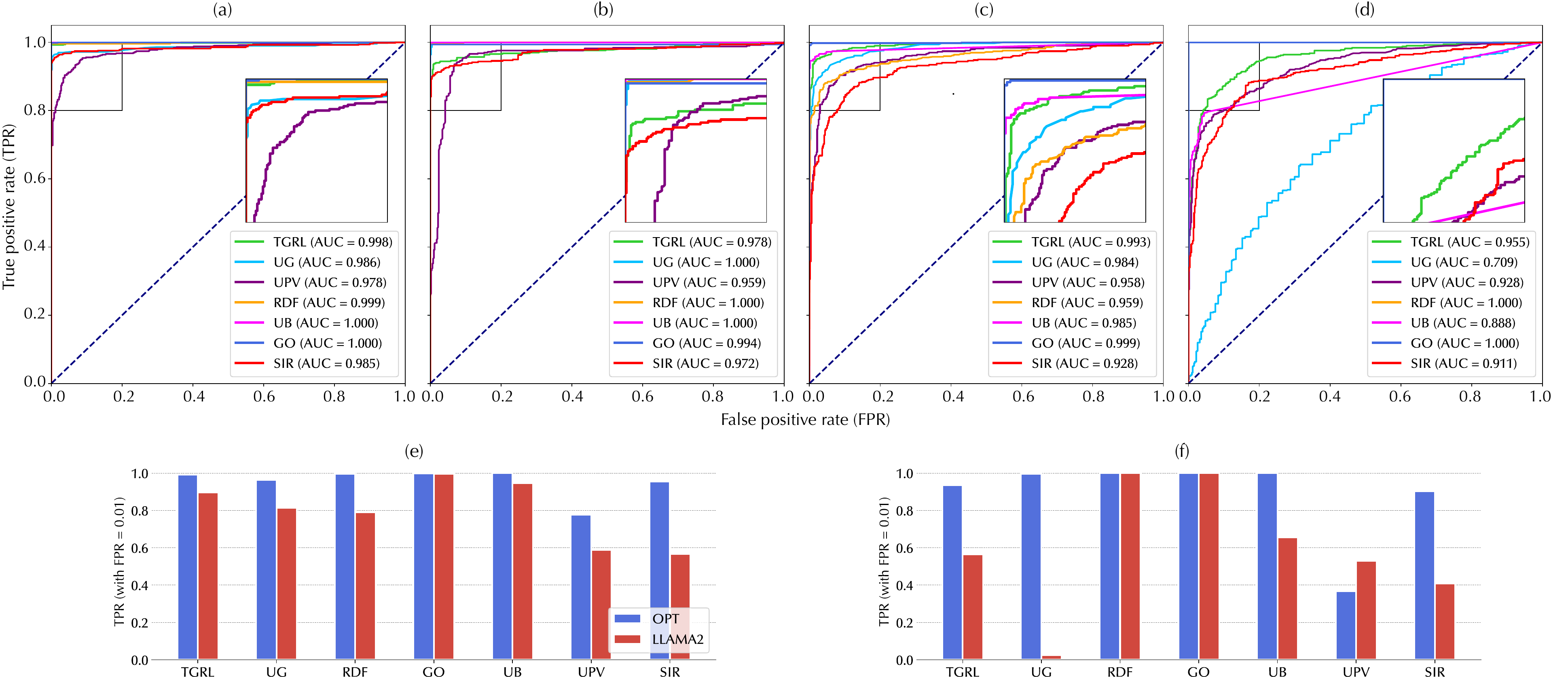}

    \caption{Overall effectiveness of different watermarkers in generating and detecting watermarked texts: ROC of (a) OPT-C4, (b) OPT-HC3, (c) Llama3-C4, and (d) Llama3-HC3; TPR (with FPR fixed as 0.01) on (e) C4 and (f) HC3. 
    \label{fig:exp_auc_clean}}
\end{figure*}

\section{Watermark Effectiveness}
\label{sec:effectiveness}

We first evaluate the effectiveness of different watermarkers. Following prior work~\citep{john,markmyword,rohith}, for each watermarker, we sample 1,000 prompts and use the LLM in combination with the watermarker to generate the watermarked texts; meanwhile, we select human responses to the same prompts as the non-watermarked texts. We then measure the accuracy of the watermarker's detector in distinguishing the watermarked and non-watermarked texts.


\subsection{Overall Effectiveness}

We measure the overall effectiveness of each method through the lens of the ROC curve. 
Figure\mref{fig:exp_auc_clean} (a-d) summarizes the overall effectiveness of existing watermarkers across different models and datasets. We have the following interesting observations.


Most watermarkers are highly effective in generating and subsequently detecting watermarked texts on OPT-1.3B as shown in Figure\mref{fig:exp_auc_clean}\,(a-b). For instance, RDF, UB, and GO all attain AUC scores above 0.99 over both C4 and HC3. Recall that C4 and HC3 represent the text completion and question-answering tasks respectively. The observation indicates that most watermarkers tend to be highly effective for relatively less capable LLMs such as OPT-1.3B, while the concrete dataset/task have a limited impact on their performance. We further validate this hypothesis under the setting of a fixed FPR. As shown in Figure\mref{fig:exp_auc_clean}\,(e-f), we fix the FPRs of all the methods to be 0.01 and measure their TPRs. Observe that all the methods achieve above 0.9 TPR, with a marginal difference across C4 and HC3.

Meanwhile, most methods observe marginal performance drops on Llama3-7B, as shown in Figure\mref{fig:exp_auc_clean}\,(c-d). For instance, compared with its performance on OPT-1.3B, the AUC of SIR drops by 0.11 and 0.08 on C4 and HC3 respectively. This observation aligns with previous research ~\citep{rohith}, indicating that watermarkers are more effective on OPT compared with Llama3. This phenomenon can be partly understood through the following explanation. In contrast of less capable LLMs (e.g., OPT-1.3B), Llama3-7B typically produces texts of lower perplexity. Since most watermarkers inject watermarks by slightly altering the next-token distribution, the lower perplexity in Llama3's outputs hampers the effectiveness of such perturbations. Moreover, it is observed that the performance of various methods varies significantly across different datasets. For instance, UG's AUC differs by 0.28 between C4 and HC3, while UB's AUC differs by 0.11. This observation is further supported by the TPR measures at fixed FPRs (fixed as 1\%), as shown in Figure\mref{fig:exp_auc_clean} (e-f). Our findings suggest that the concrete dataset/task tends to have a larger impact on watermarkers over more capable LLMs.

\subsection{Impact of Text Length}
\label{sec:length}

\begin{figure}[ht!]
    \centering
    \includegraphics[width=0.9\linewidth]{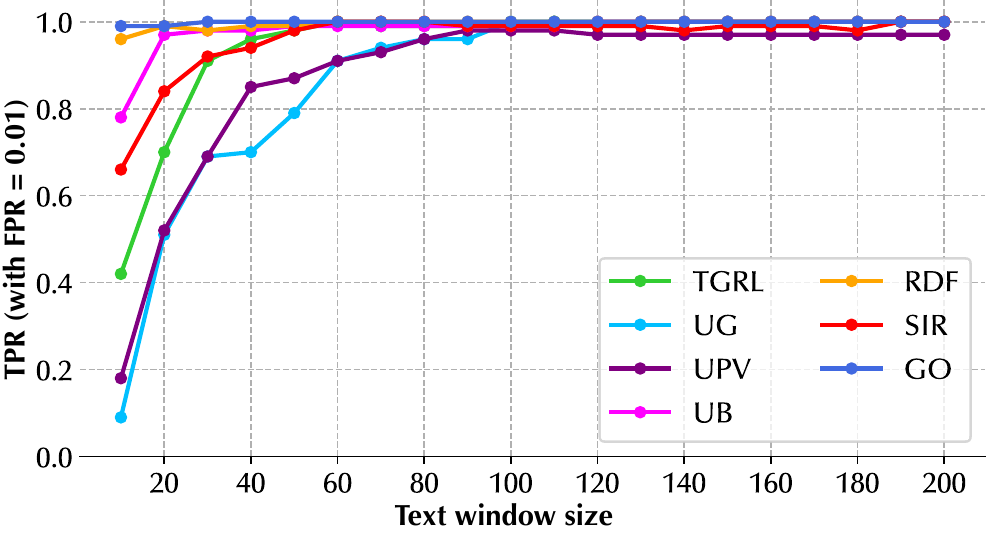}
    \caption{TPRs of watermarkers with respect to text length (with FPRs fixed as 1\%). }
    \label{fig:exp_clean_token}
\end{figure}

We evaluate how the (non-)watermarked text length (i.e., the number of tokens) impacts the performance of different methods. Specifically, we measure the TPR of each method with its FPR fixed as 1\%. In the following, we set OPT-1.3B and C4 as the default LLM and dataset. Figure\mref{fig:exp_clean_token} summarizes the results. 

Observe that as expected, the TPRs of all the methods improve as the text length grows from 1 to 200 tokens. As the text length exceeds 100 tokens, most methods reach TPRs close to 100\%. Meanwhile, different methods show varying sensitivity to the text length. For instance, GO and RDF attain 100\% TPRs with only 20 tokens, while UG reaches only around 50\% TPR under the same setting. This can be explained as follows. Both GO and RDF use distribution transform-based samplers, which, conditional on given randomness (e.g., random permutation), generate the next token deterministically. Meanwhile, other methods randomly sample the next token from a given pool (e.g., green lists). Thus, GO and RDF tend to have stronger signals per token for watermark detection. 





  

\subsection{Impact of Temperature}

The temperature $\tau$ is a key parameter that affects a watermarker's generative dynamics: intuitively, a higher $\tau$ makes the sampling over the next-token distribution $\tilde{p}(x_t| x_{< t})$ more random. Here we evaluate how the setting of $\tau$ in each watermaker's generator may impact its effectiveness. Note that, unlike other watermarkers, RDF and GO do not generate the next-token distribution $\tilde{p}(x_t| x_{< t})$ explicitly, we thus exclude them from the evaluation.


\begin{figure}[ht!]
    \setlength{\abovecaptionskip}{3pt}  
    \centering
    
    \includegraphics[width=\linewidth]{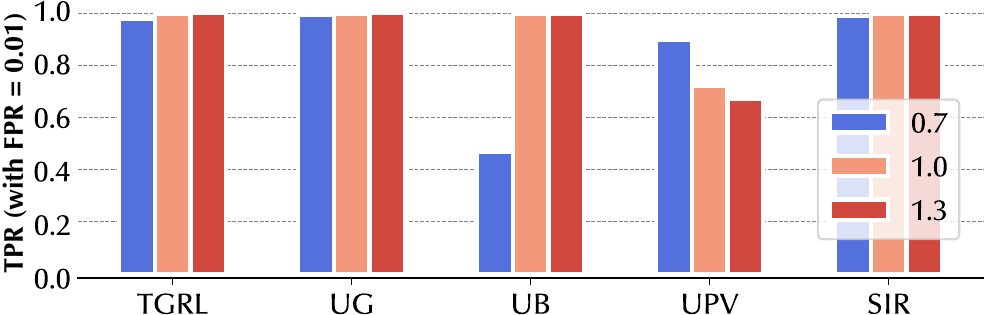}
    \caption{TPRs of watermarkers with respect to the temperature setting (with FPRs fixed as 1\%).}
    \label{fig:exp_temper_tpr}
\end{figure}

Figure\mref{fig:exp_temper_tpr} compares how the TPRs of different watermarkers vary with the setting of $\tau =$ 0.7, 1.0, and 1.3 (with FPRs fixed as 1\%). Observe that the performance of most watermarkers marginally improves with $\tau$, which corroborates prior work~\citep{markmyword}. For instance, the TPR of TGRL increases by about 0.05 as $\tau$ varies from 0.7 to 1.3. Interestingly, in contrast, the TPR of UPV decreases as $\tau$ grows. This can be explained as follows. UPV employs a neural network as the detector that depends on general textual features, which tends to be more sensitive to increasing randomness, compared with other watermarkers that rely on specific watermark signals (e.g., green/red-listed tokens). 



\section{Watermark Fidelity} 
\label{sec:fidelity}
\begin{figure}[htbp]
    \setlength{\abovecaptionskip}{3pt}  
    \centering
    \includegraphics[width=\linewidth]{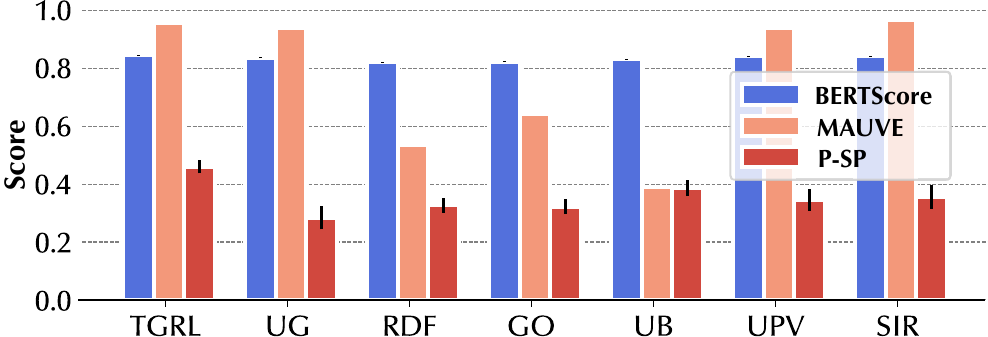}
    \caption{Fidelity preservation of different watermarkers.}
    \label{fig:exp_clean_sentiment-main}
\end{figure}

We evaluate the impact of different watermarkers on the text quality. We compare the original text $T$ and watermarked text $\tilde{T}$ using the metrics (detailed in \msec{sec:metrics}) of BERTScore~\citep{zhang2020bertscore}, P-SP~\citep{wieting2021paraphrastic}, and MAUVE~\citep{pillutla2023mauve}. The results are summarized in Figure\mref{fig:exp_clean_sentiment-main}.




i) A majority of watermarkers well preserve the semantics of original texts, as indicated by their high BERTScore and MAUVE scores. Note that the P-SP scores of all the watermarkers are relatively lower than their BERTScore and MAUVE scores. This is due to their different emphases: P-SP measures the average similarity between the tokens in $T$ and $\tilde{T}$, while BERTScore calculates the maximum similarity between the tokens in $T$ and $\tilde{T}$. ii) Meanwhile, RDF, GO, and UB are less effective in preserving the quality of original texts, which can be attributed to their additional constraints of indistinguishability: the expectation of the watermarker's next-token distribution is identical to the LLM's next-token distribution (i.e., indistinguishability). This observation suggest that there exists an inherent trade-off between the desiderata of quality and indistinguishability.

\section{Additional Results}
\label{sec:additional}

\subsection{Multi-bit Watermarking}

\begin{figure}[!ht]
    \centering
    \includegraphics[width=0.9\linewidth]{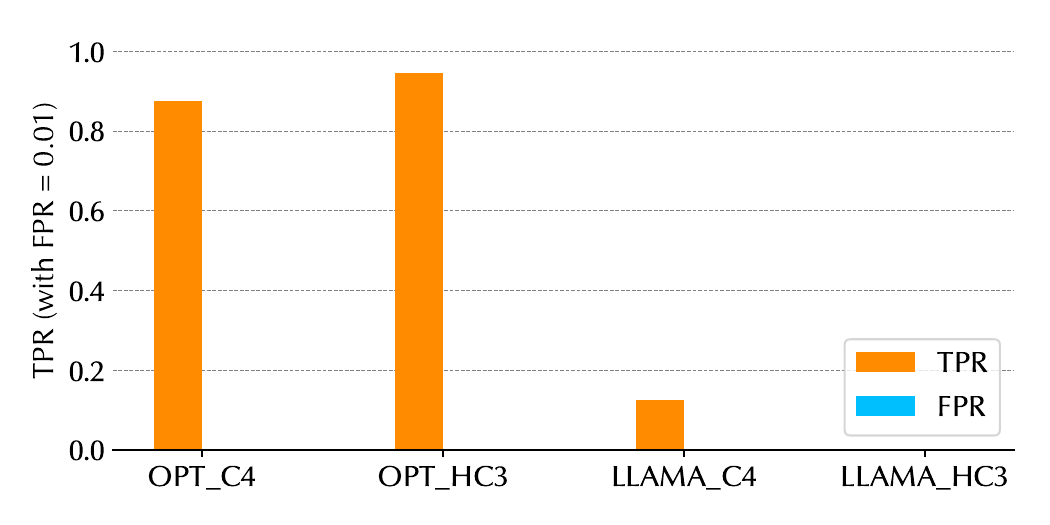}
    \caption{TPRs of CTWL (with FPRs fixed as 0.01) on different LLMs (OPT and Llama3) and datasets (C4 and HC3).\label{fig:exp_lean_clean}}
\end{figure}

\begin{figure}[!ht]
    \centering
    \includegraphics[width=0.9\linewidth]{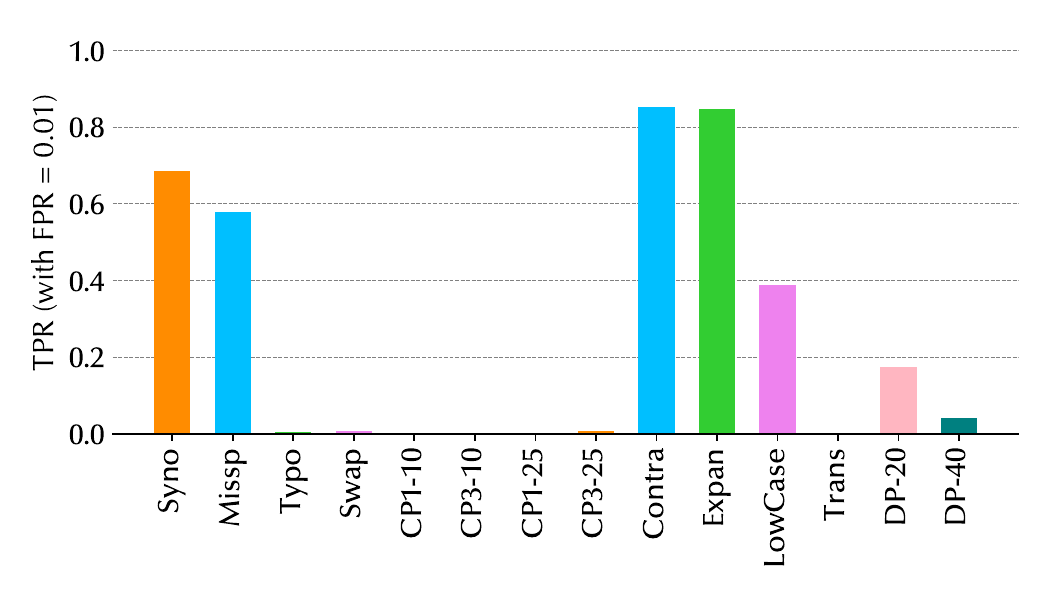}
    \caption{TPRs of CTWL (with FPRs fixed as 0.01) against various attacks.\label{fig:exp_lean_atatck}}
\end{figure}

While our study focuses on one-bit watermarkers, for completeness, we also evaluate CTWL~\citep{lean}, a multi-bit watermarker. In contrast of one-bit watermarkers that encode only a single bit of information (i.e., whether a given text is watermarked), a multi-bit watermarker can encode multiple bits of information into the watermarked text, such as the generating model, the date of generation, and other details. However, despite its larger information capacity, we find that a multi-bit watermarker is typically less robust compared to one-bit watermarkers, as illustrated in Figure\mref{fig:exp_lean_clean} and \mref{fig:exp_lean_atatck}.

The test results in the basic encoding and detection scenario highlight the sensitivity of CTWL to different language models. Although it performs well with the OPT model across both C4 and HC3 datasets, its TPR on Llama3 is extremely low, and it becomes completely ineffective when tested on Llama3 using the HC3 dataset. This is due to the lower model perplexity of Llama3. We conduct additional experiments to compare the model perplexity on the same WikiText dataset, and the results show that Llama3 has a perplexity about 6.15, while OPT has about 12.43. Lower model perplexity results to larger fluctuations in the logits produced by the model, makes it more challenging for watermark injection (e.g., increase smaller logits to exceed larger ones).

When facing the attacks, CTWL is more vulnerable than one-bit methods. It is particularly vulnerable to the copy-pasting attack, which can nearly disable the method as the TPR drops to near zero. Additionally, CTWL is highly susceptible to typoing, swapping, translating, lowercasing, and Dipper attacks, which generally do not affect many one-bit methods as severely. Despite the capacity of multi-bit methods to embed more information, their high sensitivity to language model variations and various attacks is a crucial limitation that needs to be addressed in future research.

\subsection{Additional Robustness of Watermarkers}
\label{sec:add_robust}
\textcolor{black}{Here we present the full robustness evaluation results across different models and datasets, which strongly corroborate our primary experimental findings. These experiments revealed several significant patterns that align with earlier observations, demonstrating that the relative robustness rankings of watermarking methods remain consistent regardless of model architecture or dataset characteristics. This consistency suggests that robustness is primarily determined by the inherent algorithmic properties of each method.}

\textcolor{black}{Table \mref{tbl:full_attack} shows the detailed results. Notably, UPV and SIR exhibited substantial vulnerability, with watermark detectability degrading even under simple Linguistic variation and Lexical editing attacks, indicating fundamental limitations in their robustness across model scales. Text-mixing attacks posed significant challenges for most watermarkers—particularly UPV, which failed completely under such conditions—while RDF and GO maintained reliable detection, highlighting key differences in algorithmic resilience. Furthermore, under paraphrasing attacks, both UG and RDF consistently achieved high detection rates, further supporting the conclusion that inherent watermarking design, rather than model- or dataset-specific factors, governs robustness against adversarial manipulations.}

\begin{table*}[h]
\centering
\small
\def\arraystretch{1.2}
\setlength{\tabcolsep}{1.5pt}
\definecolor{myblue}{rgb}{0.8,0.8,1}
\definecolor{myred}{rgb}{1,0.8,0.8}
\resizebox{0.98\textwidth}{!}{

\begin{tabular}{lccccccccccccccc}
    \toprule
\multirow{2}{*}{\makecell{Water\\marker}} & \multirow{2}{*}{CLEAN} & \multicolumn{3}{c}{Linguistic variation} & \multicolumn{4}{c}{Lexical editing} & \multicolumn{4}{c}{Text-mixing} & \multicolumn{3}{c}{Paraphrasing} \\

\cmidrule(lr){3-5} \cmidrule(lr){6-9} \cmidrule(lr){10-13} \cmidrule(lr){14-16}

 &  & Contra & Expan & LowCase & Swap & Typo & Syno & Missp & CP1-10 & CP3-10 & CP1-25 & CP3-25 & DP-20 & DP-40 & Trans \\ 

\midrule

\multicolumn{16}{c}{OPT + C4}\\ 
\midrule

TGRL & \cellcolor{myred!99}0.992 & \cellcolor{myred!99}0.994 & \cellcolor{myred!99}0.994 & \cellcolor{myred!96}0.984 & \cellcolor{myred!74}0.874 & \cellcolor{myred!58}0.788 & \cellcolor{myred!99}0.996 & \cellcolor{myred!99}0.992 & \cellcolor{myblue!97}0.034 & \cellcolor{myblue!83}0.176 & \cellcolor{myblue!87}0.126 & \cellcolor{myred!67}0.836 & \cellcolor{myred!90}0.952 & \cellcolor{myred!72}0.858 & \cellcolor{myred!30}0.652 \\ 
UG & \cellcolor{myred!93}0.964 & \cellcolor{myred!93}0.963 & \cellcolor{myred!93}0.965 & \cellcolor{myred!76}0.879 & \cellcolor{myred!91}0.956 & \cellcolor{myred!77}0.884 & \cellcolor{myred!93}0.965 & \cellcolor{myred!91}0.956 & \cellcolor{myblue!87}0.131 & \cellcolor{myblue!78}0.219 & \cellcolor{myblue!81}0.192 & \cellcolor{myred!7}0.534 & \cellcolor{myred!88}0.940 & \cellcolor{myred!83}0.916 & \cellcolor{myred!32}0.658 \\ 
UPV & \cellcolor{myred!55}0.776 & \cellcolor{myred!54}0.772 & \cellcolor{myred!63}0.814 & \cellcolor{myred!37}0.685 & \cellcolor{myred!33}0.667 & \cellcolor{myblue!29}0.355 & \cellcolor{myred!53}0.764 & \cellcolor{myred!39}0.695 & \cellcolor{myblue!90}0.050 & \cellcolor{myblue!85}0.073 & \cellcolor{myblue!52}0.242 & \cellcolor{myblue!10}0.450 & \cellcolor{myblue!41}0.296 & \cellcolor{myblue!89}0.057 & \cellcolor{myblue!99}0.006 \\ 
RDF & \cellcolor{myred!99}0.996 & \cellcolor{myred!99}0.993 & \cellcolor{myred!99}0.993 & \cellcolor{myred!96}0.983 & \cellcolor{myred!90}0.951 & \cellcolor{myred!92}0.961 & \cellcolor{myred!98}0.990 & \cellcolor{myred!99}0.993 & \cellcolor{myblue!81}0.094 & \cellcolor{myblue!21}0.396 & \cellcolor{myblue!14}0.430 & \cellcolor{myred!95}0.974 & \cellcolor{myred!98}0.988 & \cellcolor{myred!92}0.962 & \cellcolor{myblue!69}0.154 \\ 
SIR & \cellcolor{myred!91}0.954 & \cellcolor{myred!91}0.954 & \cellcolor{myred!91}0.954 & \cellcolor{myred!77}0.886 & \cellcolor{myred!86}0.932 & \cellcolor{myred!13}0.566 & \cellcolor{myred!87}0.936 & \cellcolor{myred!81}0.904 & \cellcolor{myblue!72}0.138 & \cellcolor{myblue!58}0.210 & \cellcolor{myblue!64}0.178 & \cellcolor{myblue!7}0.466 & \cellcolor{myred!80}0.902 & \cellcolor{myred!63}0.814 & \cellcolor{myred!64}0.818 \\ 
GO & \cellcolor{myred!100}0.998 & \cellcolor{myred!100}0.998 & \cellcolor{myred!100}0.998 & \cellcolor{myred!98}0.988 & \cellcolor{myred!84}0.920 & \cellcolor{myred!70}0.852 & \cellcolor{myred!99}0.996 & \cellcolor{myred!100}1.000 & \cellcolor{myblue!64}0.181 & \cellcolor{myred!18}0.589 & \cellcolor{myred!24}0.620 & \cellcolor{myred!98}0.992 & \cellcolor{myred!94}0.971 & \cellcolor{myred!69}0.847 & \cellcolor{myred!86}0.928 \\ 
\midrule

\multicolumn{16}{c}{LLAMA + C4}\\ 
\midrule

TGRL & \cellcolor{myred!89}0.896 & \cellcolor{myred!86}0.864 & \cellcolor{myred!89}0.892 & \cellcolor{myred!75}0.75 & \cellcolor{myred!41}0.408 & \cellcolor{myblue!66}0.34 & \cellcolor{myred!82}0.818 & \cellcolor{myred!78}0.78 & \cellcolor{myblue!99}0.014 & \cellcolor{myblue!85}0.074 & \cellcolor{myblue!82}0.088 & \cellcolor{myred!35}0.348 & \cellcolor{myred!54}0.536 & \cellcolor{myred!33}0.334 & \cellcolor{myblue!80}0.196 \\ 
UG & \cellcolor{myred!81}0.814 & \cellcolor{myred!79}0.791 & \cellcolor{myred!82}0.823 & \cellcolor{myred!70}0.697 & \cellcolor{myred!84}0.843 & \cellcolor{myblue!64}0.357 & \cellcolor{myred!75}0.75 & \cellcolor{myred!75}0.755 & \cellcolor{myblue!98}0.025 & \cellcolor{myblue!90}0.054 & \cellcolor{myblue!90}0.050 & \cellcolor{myblue!81}0.190 & \cellcolor{myred!54}0.771 & \cellcolor{myred!27}0.633 & \cellcolor{myblue!69}0.344 \\ 
UPV & \cellcolor{myred!59}0.588 & \cellcolor{myred!50}0.502 & \cellcolor{myred!60}0.598 & \cellcolor{myred!49}0.494 & \cellcolor{myred!39}0.388 & \cellcolor{myblue!96}0.018 & \cellcolor{myred!51}0.512 & \cellcolor{myred!39}0.392 & \cellcolor{myblue!99}0.005 & \cellcolor{myblue!100}0.002 & \cellcolor{myblue!100}0.002 & \cellcolor{myblue!99}0.007 & \cellcolor{myred!36}0.362 & \cellcolor{myred!35}0.355 & \cellcolor{myblue!96}0.042 \\ 
RDF & \cellcolor{myred!79}0.790 & \cellcolor{myred!74}0.735 & \cellcolor{myred!73}0.725 & \cellcolor{myred!62}0.622 & \cellcolor{myred!33}0.333 & \cellcolor{myred!89}0.888 & \cellcolor{myred!74}0.738 & \cellcolor{myred!82}0.824 & \cellcolor{myblue!98}0.021 & \cellcolor{myblue!91}0.045 & \cellcolor{myblue!91}0.045 & \cellcolor{myred!31}0.307 & \cellcolor{myred!47}0.471 & \cellcolor{myred!30}0.303 & \cellcolor{myblue!99}0.010 \\ 
SIR & \cellcolor{myred!57}0.566 & \cellcolor{myred!7}0.066 & \cellcolor{myred!8}0.078 & \cellcolor{myred!8}0.078 & \cellcolor{myblue!92}0.076 & \cellcolor{myblue!99}0.010 & \cellcolor{myblue!93}0.064 & \cellcolor{myblue!95}0.050 & \cellcolor{myblue!97}0.034 & \cellcolor{myblue!97}0.026 & \cellcolor{myblue!96}0.042 & \cellcolor{myblue!95}0.048 & \cellcolor{myred!4}0.040 & \cellcolor{myred!8}0.081 & \cellcolor{myblue!90}0.100 \\ 
GO & \cellcolor{myred!100}0.996 & \cellcolor{myred!100}0.996 & \cellcolor{myred!100}0.996 & \cellcolor{myred!96}0.959 & \cellcolor{myred!64}0.643 & \cellcolor{myred!56}0.556 & \cellcolor{myred!99}0.989 & \cellcolor{myred!99}0.987 & \cellcolor{myblue!92}0.058 & \cellcolor{myblue!77}0.228 & \cellcolor{myblue!71}0.286 & \cellcolor{myred!82}0.825 & \cellcolor{myred!90}0.900 & \cellcolor{myred!27}0.623 & \cellcolor{myred!81}0.814 \\

\midrule
\multicolumn{16}{c}{QWEN + Paper Conclusion}\\ 
\midrule

TGRL & \cellcolor{myred!99}0.993 & \cellcolor{myred!94}0.944 & \cellcolor{myred!94}0.944 & \cellcolor{myred!83}0.831 & \cellcolor{myblue!14}0.429 & \cellcolor{myblue!44}0.222 & \cellcolor{myred!77}0.887 & \cellcolor{myred!82}0.915 & \cellcolor{myblue!100}0.000 & \cellcolor{myblue!67}0.667 & \cellcolor{myred!67}0.833 & \cellcolor{myred!67}0.833 & \cellcolor{myblue!67}0.667 & \cellcolor{myblue!30}0.485 & \cellcolor{myblue!44}0.222 \\ 
UG & \cellcolor{myred!99}0.993 & \cellcolor{myred!86}0.857 & \cellcolor{myred!83}0.833 & \cellcolor{myred!98}0.976 & \cellcolor{myred!93}0.929 & \cellcolor{myred!93}0.930 & \cellcolor{myred!98}0.976 & \cellcolor{myred!62}0.738 & \cellcolor{myred!86}0.857 & \cellcolor{myred!71}0.714 & \cellcolor{myred!71}0.714 & \cellcolor{myred!99}0.991 & \cellcolor{myred!98}0.976 & \cellcolor{myred!88}0.877 & \cellcolor{myred!92}0.921 \\ 
UPV & \cellcolor{myblue!20}0.400 & \cellcolor{myblue!20}0.400 & \cellcolor{myblue!20}0.400 & \cellcolor{myblue!14}0.430 & \cellcolor{myblue!84}0.080 & \cellcolor{myblue!100}0.000 & \cellcolor{myblue!24}0.380 & \cellcolor{myblue!28}0.360 & \cellcolor{myblue!100}0.000 & \cellcolor{myblue!100}0.000 & \cellcolor{myblue!100}0.000 & \cellcolor{myblue!100}0.000 & \cellcolor{myblue!60}0.200 & \cellcolor{myblue!84}0.080 & \cellcolor{myblue!52}0.240 \\ 
RDF & \cellcolor{myred!100}0.999 & \cellcolor{myred!100}0.998 & \cellcolor{myred!100}0.996 & \cellcolor{myred!100}0.996 & \cellcolor{myred!98}0.976 & \cellcolor{myred!98}0.979 & \cellcolor{myred!99}0.991 & \cellcolor{myred!99}0.993 & \cellcolor{myred!87}0.872 & \cellcolor{myred!89}0.893 & \cellcolor{myred!93}0.932 & \cellcolor{myred!98}0.978 & \cellcolor{myred!91}0.905 & \cellcolor{myred!69}0.738 & \cellcolor{myred!98}0.978 \\ 
UB & \cellcolor{myred!97}0.980 & \cellcolor{myred!100}1.000 & \cellcolor{myred!100}1.000 & \cellcolor{myred!96}0.962 & \cellcolor{myblue!100}0.000 & \cellcolor{myblue!96}0.033 & \cellcolor{myred!92}0.921 & \cellcolor{myred!100}1.000 & \cellcolor{myblue!98}0.018 & \cellcolor{myblue!95}0.042 & \cellcolor{myblue!100}0.000 & \cellcolor{myblue!51}0.485 & \cellcolor{myred!51}0.514 & \cellcolor{myblue!89}0.103 & \cellcolor{myblue!74}0.255 \\
SIR & \cellcolor{myred!98}0.978 & \cellcolor{myblue!58}0.580 & \cellcolor{myblue!50}0.500 & \cellcolor{myblue!40}0.420 & \cellcolor{myblue!36}0.320 & \cellcolor{myred!72}0.320 & \cellcolor{myblue!44}0.440 & \cellcolor{myblue!56}0.560 & \cellcolor{myblue!32}0.340 & \cellcolor{myblue!36}0.360 & \cellcolor{myblue!36}0.360 & \cellcolor{myblue!46}0.460 & \cellcolor{myblue!40}0.400 & \cellcolor{myblue!60}0.300 & \cellcolor{myblue!100}0.000 \\ 
GO & \cellcolor{myred!100}0.996 & \cellcolor{myred!100}0.996 & \cellcolor{myred!100}0.996 & \cellcolor{myred!99}0.994 & \cellcolor{myred!98}0.982 & \cellcolor{myred!96}0.956 & \cellcolor{myred!99}0.986 & \cellcolor{myred!100}0.996 & \cellcolor{myred!86}0.864 & \cellcolor{myred!89}0.887 & \cellcolor{myred!95}0.946 & \cellcolor{myred!98}0.982 & \cellcolor{myblue!60}0.640 & \cellcolor{myblue!56}0.560 & \cellcolor{myblue!67}0.667 \\

\bottomrule
\end{tabular}}
\caption{\small \textcolor{black}{Attack resilience of LLM watermarkers. The intensity of red shading indicates higher values, while the intensity of blue shading indicates lower values, with 0.5 serving as the threshold between the two color gradients. \label{tbl:full_attack}}}
\end{table*}

\subsection{Different Styles Datasets}
\label{sec:add_dataset}
\textcolor{black}{To comprehensively evaluate watermarking performance across different text generation scenarios, we created diverse datasets spanning multiple styles and domains. The Paper Conclusion dataset was constructed by collecting research papers from arXiv and having LLMs summarize their key findings. We also incorporated Law Stack Exchange, a legal domain question-answering dataset sourced from the Stack Exchange forum, and WritingPrompts, a creative writing dataset focused on story completion tasks.
These datasets can be categorized into three distinct styles: completion-style (C4 and Story Completion datasets\mcite{writing_prompts}), question-answering (HC3 and Law Stack Exchange datasets\mcite{law_stack_exchange}), and summarization (Paper Conclusion dataset). Our analysis revealed several significant findings.  Table \mref{tbl:diff_dataset} shows the results. }

\textcolor{black}{First, all watermarkers demonstrated notably lower performance on question-answering style datasets. This degradation was particularly pronounced for UPV and SIR, which achieved only around 0.5 TPR on these datasets. This can be attributed to the prevalence of domain-specific terminology and fixed expressions in Q\&A datasets, which inherently constrain the model's linguistic choices and make watermark embedding more challenging.
Second, robust watermarkers like RDF and GO maintained exceptional performance across all dataset styles, demonstrating their versatility and reliability. RDF achieved consistently high TPR scores ranging from 0.934 to 0.999, while GO demonstrated even more stable performance with TPR values between 0.978 and 0.998. These results highlight an important pattern: watermarkers that demonstrate strong baseline performance maintain their effectiveness across different text generation styles, while less stable or lower-performing watermarkers are particularly vulnerable to degradation when applied to question-answering formats.}
\textcolor{black}{Notably, the high watermarking effectiveness on Paper Conclusion tasks (TPR > 0.99 for most methods) can be attributed to the inherent flexibility in academic summarization. Unlike Q\&A tasks that require specific terminology and fixed expressions, paper conclusions allow for more freedom in expression while conveying the same core findings. This flexibility in language choice and sentence construction provides more opportunities for watermark embedding without compromising the semantic accuracy of the summary. }

\begin{table}[h]
\centering
\small
\def\arraystretch{1.3}
\setlength{\tabcolsep}{1pt}
\definecolor{myblue}{rgb}{0.8,0.8,1}
\definecolor{myred}{rgb}{1,0.8,0.8}
\resizebox{0.98\linewidth}{!}{

\begin{tabular}{lccccc}
    \toprule

\multirow{2}{*}{\makecell{Water\\marker}} &\multicolumn{1}{c}{Summarization} & \multicolumn{2}{c}{Completion} &\multicolumn{2}{c}{Question-Answer}\\

\cmidrule(lr){2-2} \cmidrule(lr){3-4} \cmidrule(lr){5-6}

& Conclude & C4 & StoryComple & HC3 & LawQA \\ 
\midrule
TGRL & \cellcolor{myred!99}0.993 & \cellcolor{myred!98}0.992 & \cellcolor{myred!98}0.992 & \cellcolor{myred!85}0.926 & \cellcolor{myred!87}0.934 \\ 
UG & \cellcolor{myred!99}0.993 & \cellcolor{myred!94}0.972 & \cellcolor{myred!96}0.981 & \cellcolor{myred!73}0.864 & \cellcolor{myred!78}0.891 \\ 
UPV & \cellcolor{myred!80}0.400 & \cellcolor{myblue!80}0.540 & \cellcolor{myred!12}0.560 & \cellcolor{myblue!80}0.460 & \cellcolor{myblue!80}0.430 \\ 
RDF & \cellcolor{myred!100}0.999 & \cellcolor{myred!100}0.998 & \cellcolor{myred!99}0.996 & \cellcolor{myred!89}0.945 & \cellcolor{myred!87}0.934 \\ 
SIR & \cellcolor{myred!96}0.978 & \cellcolor{myred!93}0.965 & \cellcolor{myred!95}0.976 & \cellcolor{myred!13}0.566 & \cellcolor{myred!26}0.631 \\ 
GO & \cellcolor{myred!99}0.996 & \cellcolor{myred!100}0.998 & \cellcolor{myred!99}0.994 & \cellcolor{myred!97}0.986 & \cellcolor{myred!96}0.978 \\ 
\bottomrule
\end{tabular}
}
\caption{\small \textcolor{black}{Performance of watermarkers across different datasets on Qwen2.5-14B.}}
\label{tbl:diff_dataset}
\end{table}

\subsection{Fidelity Preservation of attacks}
\label{sec:appendix_sentiment}

\begin{figure}[h]
    \centering
    \includegraphics[width=\linewidth]{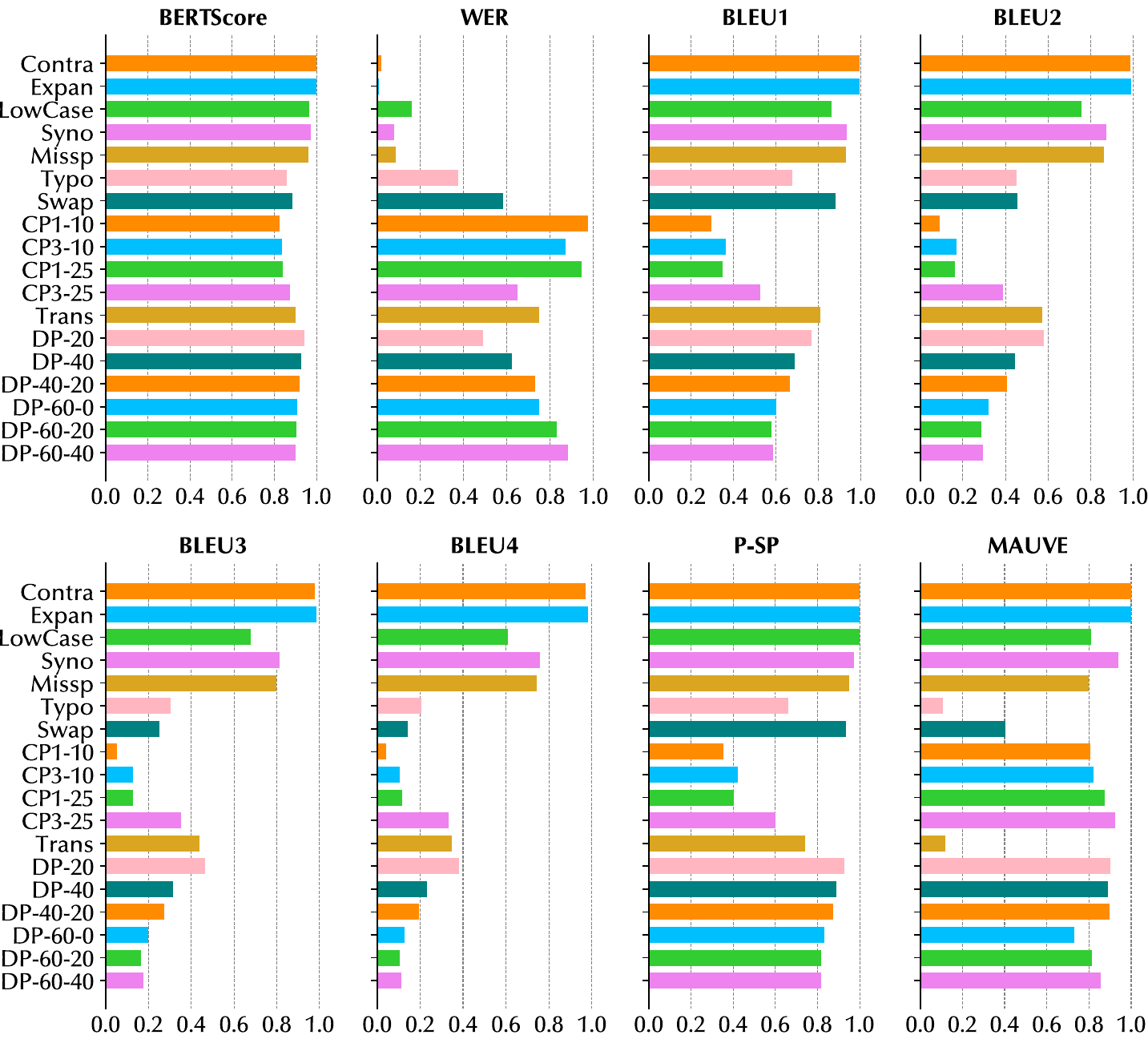}
    \caption{Quality preservation of different attacks on GO.}
    \label{fig:attack_sentiment_diff_full}
\end{figure}

Figure\mref{fig:attack_sentiment_diff_full} illustrates the quality preservation of different attacks on GO, with similar results on other watermarkers,

\section{Additional Related Work}
\label{sec:related}

We survey the relevant literature in the following categories: i) detection of LLM-generated texts, ii) LLM watermarking, iii) attacks on LLM watermarking, and iv) evaluation of LLM watermarkers.


{\bf Detection of LLM-generated texts.}
The advances in LLMs also give rise to their possible misuses. There is thus a pressing need for the capability of distinguishing LLM- and human-generated texts. Initial work attempts to either train classifiers using LLM- and human-generated texts\mcite{detect_gpt} or to leverage intrinsic characteristics of LLM-generated texts (e.g., perplexity and variability in length, complexity, and information density)\mcite{gptzero}. Yet, with LLMs becoming increasingly capable, the difference between LLM- and human-generated texts is narrowing, making such approaches less effective.

{\bf LLM watermarking} In response, LLM watermarking emerges as a promising alternative, which instruments the LLM generative process with statistical signals that can be subsequently detected. The existing LLM watermarking techniques can be categorized based on the stages in which they are applied\mcite{liu2023survey}: i) training-time watermarking\mcite{liu2023watermarking,tang2023did,sun2022coprotector,sun2023codemark,xu2024learning}, ii) watermarking during logit-generation\mcite{john,xuandong,xiaoniu,aiweia,aiweib,fairoze2023publicly,ren2023robust,fernandez2023three,lean,yoo2023advancing,lee2023wrote,giboulot2024watermax}, iii) watermarking during token-sampling \mcite{rohith,scott,hou2023semstamp}, and iv) post-generation watermarking\mcite{remarkllm,kiyoon,yang2023watermarking,munyer2023deeptextmark,sato2023embarrassingly,abdelnabi2021adversarial}. This study mainly focuses on training-free, pre-generation watermarking, which applies to any given LLMs and provides flexible control over multiple criteria (e.g., quality, effectiveness, and robustness). Some watermarks also consider asymmetric schemes~\citep{public-detectable-watermark}, while our study mainly focuses on symmetric watermarking schemes due to their widespread adoption.


 



The primary focus of the previous studies is to distinguish between human-written text and text generated by LLMs. In this paper, we assume that the attacker lacks access to non-watermarked text produced by recent LLM. Instead, the attacker can only utilize other LLMs to generate text that mimics human responses and get the watermarked texts from recent LLM. The generic detector, which is within the adversary's capability, is specifically designed to distinguish between watermarked and non-watermarked texts(e.g., other LLMs’ texts or human writing texts).

{\bf Attacks on LLM watermarking.}
One critical property of an LLM watermarker is its robustness against potential attacks. A variety of attacks can be applied to LLM watermarking, ranging from removing the embedded watermark to uncovering the green/red lists. For instance, Dipper\mcite{retrieval-defense} is a widely used paraphrasing attack to evaluate the robustness of LLM watermarkers \mcite{john,xuandong} against watermark removal attacks; the watermark stealing attack\mcite{jovanovic2024watermark} is proposed to identify green-list tokens and to replace them with red-list tokens, targeting TGRL\mcite{john} and UG\mcite{xuandong} to further launch spoofing or removal attacks. This study primarily focuses on watermark removal attacks as they can target any LLM watermarkers and have profound implications in practice (e.g., disinformation, academic cheating, and automated phishing).

{\bf Evaluation of LLM watermarkers.}
As LLM watermarkers become prevalent, recent work attempts to benchmark the performance of various watermarkers. However, current studies either focus solely on the effectiveness of watermarking or have limited assessments of robustness. For instance, WaterBench\mcite{waterbench} compares the effectiveness of TGRL\mcite{john} and UG\mcite{xuandong} under varying hyper-parameter settings (e.g., prompt length); LLM-judger\mcite{singh2023new} uses GPT-3.5-Turbo as a judge to evaluate the effectiveness of RDF\mcite{rohith} and TGRL\mcite{john} and employs a binary classifier based on MLP to distinguish between watermarked and non-watermarked texts; MarkMyWords\mcite{markmyword} compares the effectiveness of four watermarkers including TGRL\mcite{john}, GO\mcite{scott}, RDF\mcite{rohith}, and UW\mcite{undetetable}, and evaluates the robustness of TGRL against watermark removal attacks.

\textcolor{black}{While our study shares some evaluation components with MarkMyWords, our contributions differ substantially in scope and depth. MarkMyWords includes robustness testing as part of a broader evaluation, whereas our work provides a dedicated and systematic framework specifically for stress-testing watermark robustness. This specialized focus enables deeper exploration of attack types, adaptive strategies, and evaluation metrics. Methodologically, we introduce several new dimensions not covered in prior work, such as causal analysis of watermark design choices, comparisons between specific and generic detectors, and adaptive attacks mounted by adversaries using generic detectors. Furthermore, we establish a standardized benchmark comprising 12 representative attacks and 8 evaluation metrics—facilitating fine-grained robustness analysis, including next-token distribution comparisons and what-if attack simulations.}

\textcolor{black}{Our systematic approach has uncovered several novel findings that go beyond the scope of existing evaluations—for example, challenging assumptions in prior work (see Table~\ref{tbl:compare}) and identifying key trade-offs in design robustness. We acknowledge the need to better articulate these distinctions and will revise the related work section to more clearly position our contributions relative to prior efforts such as MarkMyWords.}

To our best knowledge, this is the first study solely dedicated to evaluating the robustness of LLM watermarkers against removal attacks. Our goal is to understand how design decisions influence resistance to adversarial perturbations and to establish best practices for deploying watermarkers in real-world adversarial environments.

{\bf Broader LLM and Security Research}
A growing body of work explores how to make LLM training more efficient and adaptable. Communication-efficient federated fine-tuning~\cite{liu2025ecoloracommunicationefficientfederatedfinetuning}, sequential recommendation frameworks~\cite{liu2024sequentialllmframeworkfashion}, and distributed training systems such as MalleTrain~\cite{10.1145/3629526.3645035} address scalability and resource utilization. Related efforts include FedCust for hyperparameter customization~\cite{zawad2025fedcust}, serverless orchestration for cloud-scale learning~\cite{ali2025enabling,ma2024towards}, and pruning or tuning strategies for domain-specific LLMs~\cite{lu2024all}. Complementary advances in graph-augmented learning for medical tasks~\cite{wang2024self} and classical work on efficient model execution~\cite{wu2020intermittent,wu2020enabling} further highlight the importance of scalable and resource-aware deployment.  

In parallel, extensive research has been devoted to improving LLM safety and robustness. This includes analyzing memorization through dynamic soft prompting~\cite{wang2024unlocking}, investigating systemic vulnerabilities in GraphRAG~\cite{liang2025graphrag}, and revisiting classical model extraction attacks~\cite{liang2024model}. Defenses have been proposed through KV eviction against jailbreaks~\cite{jiang2024robustkv}, clean-data curation~\cite{liu2024robustifying,liu2025datadefenserolecuration}, and dynamic token reweighting for vision–language models~\cite{jiang2025robustifyingvisionlanguagemodelsdynamic}. Other directions include weak-to-strong jailbreak transfer~\cite{liang2025autoranweaktostrongjailbreakinglarge}, self-improving steering for alignment~\cite{zhu2025selfimprovingmodelsteering}, and even designing self-destructive mechanisms for LLMs~\cite{wang2025selfdestructivelanguagemodel}. At the same time, works have highlighted broader risks and applications, such as privacy leakage from explainability~\cite{10646875}, vulnerabilities in moderation guardrails~\cite{zhuang2025exploring}, and LLMs as copilots in cyber defense~\cite{liu2025cyber}.  

Finally, cross-domain studies enrich the LLM security landscape by offering conceptual parallels. Research on quantum optimization~\cite{zhuang2025largelanguagemodelshelp}, digital forensics~\cite{yin2025digitalforensicsagelarge}, and multimodal prognostics~\cite{Wangjqt2024,wangtase2025,wangtim2024,wangcase2023} illustrates how adaptive modeling, reliability analysis, and self-monitoring are critical across domains. These perspectives provide additional context for understanding LLM vulnerabilities and defenses, underscoring that robustness and trustworthiness are shared challenges spanning machine learning, distributed systems, and safety-critical applications.

\section{Evaluation Guidelines for Future LLM Watermarking Research}
\label{sec:eva_guide}
Next, we propose a set of guidelines for evaluating the robustness of LLM watermarkers. These guidelines incorporate our findings in \msec{sec:eval} and \msec{sec:discussion}, providing a minimal checklist to claim the robustness of an LLM watermarker.

{\bf LLMs and tasks.} Our experiments show that the referenced watermarkers show varying robustness across different LLMs and datasets. We speculate that there exists an intricate interplay between the watermarking mechanism, the LLM's capability, and the task's complexity. We thus recommend experimenting on i) LLMs with varying capability (e.g., as measured by perplexity), ii) datasets for different tasks (e.g., summarization and question-answering), and iii) their combinations.

{\bf Attacks.} Notably, using highly capable LLMs or applying computationally expensive rewriting can easily generate highly-quality, non-watermarked texts to evade detection; however, such attacks negate the need for watermark removal attacks in the first place. We thus recommend focusing on computationally efficient attacks such as linguistic variation, lexical editing, and lightweight paraphrasing, as well as their combinations, which reflects the risks of watermark removal attacks in practical settings.

{\bf Robustness.} It is often critical to properly set the decision threshold for a watermarker (and also the attacks) to fully assess its robustness\mcite{sok-image}, which unfortunately is often missing in the original papers. To overcome this issue, we recommend i) measuring the overall effectiveness (TPR) in terms of ROC (across different threshold settings), ii) measuring the TPR under a fixed FPR (e.g., 0.01), and iii) considering varying attack intensity.

{\bf Fidelity.} It is notoriously challenging to meaningfully measure the quality of text data\mcite{pillutla2023mauve}. We recommend employing a variety of metrics (e.g., BERTScore, P-SP, MAUVE) to comprehensively measure the quality retention of watermarkers as well as attacks. In addition, one may also leverage external tools (e.g., more advanced LLMs such as GPT-4) or human evaluation to provide a more accurate assessment if feasible.

\section{Deployment Guidelines for Adversarial Environments}
\label{sec:dev_guide}

\textcolor{black}{To complement our empirical evaluation, we distill our findings into actionable deployment guidelines for practitioners operating watermarkers in adversarial settings.}

\textcolor{black}{{\bf Watermarker selection.} Our results show that watermarkers differ significantly in their robustness to specific attack types. We recommend selecting watermarkers based on the anticipated threat landscape: RDF and UG are especially resilient against paraphrasing and text-mixing, while TGRL performs well against linguistic variations. Practitioners should avoid UPV and SIR in high-adversity settings due to their instability and vulnerability under attack combinations.}

\textcolor{black}{{\bf Detector configuration.} For watermarkers with unstable detectors (e.g., UPV), using hybrid detection schemes—combining score-based and model-based detectors or incorporating a generic detector—can mitigate failure modes. Our results (Table~\mref{tbl:generic_eva_acc}) suggest that generic detectors can complement specific ones, enhancing recall without significant degradation in precision.}

\textcolor{black}{{\bf Multi-layered defense.} Since individual attacks (e.g., typoing) are often insufficient to remove watermarks, while combined or adaptive attacks (e.g., swapping + synonymizing or GBDA) are far more effective, we recommend adopting multi-layered detection pipelines. These may involve edit-distance-aware scoring (e.g., RDF’s edit score) and robustness testing against common perturbation chains.}

\textcolor{black}{{\bf Risk assessment.} To assess real-world risk, practitioners should consider not only overall TPR/AUC, but also worst-case TPR under fixed low FPR (e.g., 1\%) across combined and high-intensity attacks. Our findings highlight that even robust watermarkers may fail under iterated paraphrasing (Figure~\mref{fig:exp_attack_gpt}).}

\textcolor{black}{{\bf Resource-aware deployment.} For cost-sensitive applications (e.g., online moderation), we recommend avoiding watermarkers requiring expensive generation/detection steps (e.g., deep model-assisted detection in UPV), and instead favoring methods like UG or RDF that are both robust and lightweight.}

%% file: main.bbl
\begin{thebibliography}{93}
\providecommand{\natexlab}[1]{#1}

\bibitem[{Aaronson and Kirchner()}]{scott}
Scott Aaronson and Hendrik Kirchner.
\newblock Watermarking gpt outputs.
\newblock \url{https:// www.scottaaronson.com/talks/watermark.ppt}.

\bibitem[{Abdelnabi and Fritz(2021)}]{abdelnabi2021adversarial}
Sahar Abdelnabi and Mario Fritz. 2021.
\newblock Adversarial watermarking transformer: Towards tracing text provenance with data hiding.
\newblock In \emph{2021 IEEE Symposium on Security and Privacy (SP)}, pages 121--140. IEEE.

\bibitem[{Ali et~al.(2025)Ali, Ma, Zawad, Aditya, Akkus, Chen, Yang, and Yan}]{ali2025enabling}
Ahsan Ali, Xiaolong Ma, Syed Zawad, Paarijaat Aditya, Istemi~Ekin Akkus, Ruichuan Chen, Lei Yang, and Feng Yan. 2025.
\newblock Enabling scalable and adaptive machine learning training via serverless computing on public cloud.
\newblock \emph{Performance Evaluation}, 167:102451.

\bibitem[{Bender et~al.(2021)Bender, Gebru, McMillan-Major, and Shmitchell}]{10.1145/3442188.3445922}
Emily~M. Bender, Timnit Gebru, Angelina McMillan-Major, and Shmargaret Shmitchell. 2021.
\newblock On the dangers of stochastic parrots: Can language models be too big?
\newblock In \emph{Proceedings of the ACM Conference on Fairness, Accountability, and Transparency (FAccT)}.

\bibitem[{Christ et~al.(2023)Christ, Gunn, and Zamir}]{undetetable}
Miranda Christ, Sam Gunn, and Or~Zamir. 2023.
\newblock Undetectable watermarks for language models.
\newblock Cryptology ePrint Archive, Paper 2023/763.

\bibitem[{Communication et~al.(2023)Communication, Barrault, Chung, Meglioli, Dale, Dong, Duquenne, Elsahar, Gong, Heffernan, Hoffman, Klaiber, Li, Licht, Maillard, Rakotoarison, Sadagopan, Wenzek, Ye, Akula, Chen, Hachem, Ellis, Gonzalez, Haaheim, Hansanti, Howes, Huang, Hwang, Inaguma, Jain, Kalbassi, Kallet, Kulikov, Lam, Li, Ma, Mavlyutov, Peloquin, Ramadan, Ramakrishnan, Sun, Tran, Tran, Tufanov, Vogeti, Wood, Yang, Yu, Andrews, Balioglu, Costa-jussà, Celebi, Elbayad, Gao, Guzmán, Kao, Lee, Mourachko, Pino, Popuri, Ropers, Saleem, Schwenk, Tomasello, Wang, Wang, and Wang}]{seamlessm4t}
Seamless Communication, Loïc Barrault, Yu-An Chung, Mariano~Cora Meglioli, David Dale, Ning Dong, Paul-Ambroise Duquenne, Hady Elsahar, Hongyu Gong, Kevin Heffernan, John Hoffman, Christopher Klaiber, Pengwei Li, Daniel Licht, Jean Maillard, Alice Rakotoarison, Kaushik~Ram Sadagopan, Guillaume Wenzek, Ethan Ye, Bapi Akula, Peng-Jen Chen, Naji~El Hachem, Brian Ellis, Gabriel~Mejia Gonzalez, Justin Haaheim, Prangthip Hansanti, Russ Howes, Bernie Huang, Min-Jae Hwang, Hirofumi Inaguma, Somya Jain, Elahe Kalbassi, Amanda Kallet, Ilia Kulikov, Janice Lam, Daniel Li, Xutai Ma, Ruslan Mavlyutov, Benjamin Peloquin, Mohamed Ramadan, Abinesh Ramakrishnan, Anna Sun, Kevin Tran, Tuan Tran, Igor Tufanov, Vish Vogeti, Carleigh Wood, Yilin Yang, Bokai Yu, Pierre Andrews, Can Balioglu, Marta~R. Costa-jussà, Onur Celebi, Maha Elbayad, Cynthia Gao, Francisco Guzmán, Justine Kao, Ann Lee, Alexandre Mourachko, Juan Pino, Sravya Popuri, Christophe Ropers, Safiyyah Saleem, Holger Schwenk, Paden Tomasello, Changhan Wang, Jeff
  Wang, and Skyler Wang. 2023.
\newblock \href {https://arxiv.org/abs/arXiv:2308.11596} {Seamlessm4t: Massively multilingual \& multimodal machine translation}.

\bibitem[{Compilatio()}]{academic-gpt}
Compilatio.
\newblock Cheating in the age of chatgpt: findings and solutions for preserving academic integrity.
\newblock \url{https://www.compilatio.net/en/blog/cheating-chatgpt}.

\bibitem[{Damodaran(2021)}]{t5}
Prithiviraj Damodaran. 2021.
\newblock Parrot: Paraphrase generation for {NLU}.

\bibitem[{Dathathri et~al.(2024)Dathathri, See, Ghaisas, Huang, McAdam, Welbl, Bachani, Kaskasoli, Stanforth, Matejovicova et~al.}]{dathathri2024scalable}
Sumanth Dathathri, Abigail See, Sumedh Ghaisas, Po-Sen Huang, Rob McAdam, Johannes Welbl, Vandana Bachani, Alex Kaskasoli, Robert Stanforth, Tatiana Matejovicova, et~al. 2024.
\newblock Scalable watermarking for identifying large language model outputs.
\newblock \emph{Nature}, 634(8035):818--823.

\bibitem[{Devlin et~al.(2019)Devlin, Chang, Lee, and Toutanova}]{bert}
Jacoband Devlin, Ming-Weiand Chang, Kentonand Lee, and Kristina Toutanova. 2019.
\newblock {BERT}: Pre-training of deep bidirectional transformers for language understanding.
\newblock In \emph{Proceedings of the Annual Meeting of the Association for Computational Linguistics (ACL)}.

\bibitem[{Di~Wang and Pan(2024)}]{Wangjqt2024}
Yuhui~Wang Di~Wang and Ershun Pan. 2024.
\newblock \href {https://doi.org/10.1080/00224065.2024.2315085} {Multimodal recognition and prognostics based on features extracted via multisensor degradation modeling}.
\newblock \emph{Journal of Quality Technology}, 56(3):244--256.

\bibitem[{Euclaise()}]{writing_prompts}
Euclaise.
\newblock Writingprompts dataset.
\newblock \url{https://huggingface.co/datasets/euclaise/writingprompts}.
\newblock Accessed: 2025-01-29.

\bibitem[{{Fairoze} et~al.(2023){Fairoze}, {Garg}, {Jha}, {Mahloujifar}, {Mahmoody}, and {Wang}}]{public-detectable-watermark}
Jaiden {Fairoze}, Sanjam {Garg}, Somesh {Jha}, Saeed {Mahloujifar}, Mohammad {Mahmoody}, and Mingyuan {Wang}. 2023.
\newblock Publicly detectable watermarking for language models.
\newblock \emph{ArXiv e-prints}.

\bibitem[{Fairoze et~al.(2023)Fairoze, Garg, Jha, Mahloujifar, Mahmoody, and Wang}]{fairoze2023publicly}
Jaiden Fairoze, Sanjam Garg, Somesh Jha, Saeed Mahloujifar, Mohammad Mahmoody, and Mingyuan Wang. 2023.
\newblock Publicly detectable watermarking for language models.
\newblock \emph{arXiv preprint arXiv:2310.18491}.

\bibitem[{Fernandez et~al.(2023)Fernandez, Chaffin, Tit, Chappelier, and Furon}]{fernandez2023three}
Pierre Fernandez, Antoine Chaffin, Karim Tit, Vivien Chappelier, and Teddy Furon. 2023.
\newblock Three bricks to consolidate watermarks for large language models.
\newblock In \emph{2023 IEEE International Workshop on Information Forensics and Security (WIFS)}, pages 1--6. IEEE.

\bibitem[{Giboulot and Teddy(2024)}]{giboulot2024watermax}
Eva Giboulot and Furon Teddy. 2024.
\newblock Watermax: breaking the llm watermark detectability-robustness-quality trade-off.
\newblock \emph{arXiv preprint arXiv:2403.04808}.

\bibitem[{Guo et~al.(2021)Guo, Sablayrolles, Jégou, and Kiela}]{gradient_based_attack}
Chuan Guo, Alexandre Sablayrolles, Hervé Jégou, and Douwe Kiela. 2021.
\newblock Gradient-based adversarial attacks against text transformers.
\newblock \emph{ArXiv e-prints}.

\bibitem[{Hou et~al.(2023)Hou, Zhang, He, Wang, Chuang, Wang, Shen, Van~Durme, Khashabi, and Tsvetkov}]{hou2023semstamp}
Abe~Bohan Hou, Jingyu Zhang, Tianxing He, Yichen Wang, Yung-Sung Chuang, Hongwei Wang, Lingfeng Shen, Benjamin Van~Durme, Daniel Khashabi, and Yulia Tsvetkov. 2023.
\newblock Semstamp: A semantic watermark with paraphrastic robustness for text generation.
\newblock \emph{arXiv preprint arXiv:2310.03991}.

\bibitem[{Hu et~al.(2024)Hu, Chen, Wu, Wu, Zhang, and Huang}]{xiaoniu}
Zhengmian Hu, Lichang Chen, Xidong Wu, Yihan Wu, Hongyang Zhang, and Heng Huang. 2024.
\newblock Unbiased watermark for large language models.
\newblock In \emph{Proceedings of the International Conference on Learning Representations (ICLR)}.

\bibitem[{Izacard et~al.(2022)Izacard, Lewis, Lomeli, Hosseini, Petroni, Schick, Dwivedi-Yu, Joulin, Riedel, and Grave}]{izacard2022atlas}
Gautier Izacard, Patrick Lewis, Maria Lomeli, Lucas Hosseini, Fabio Petroni, Timo Schick, Jane Dwivedi-Yu, Armand Joulin, Sebastian Riedel, and Edouard Grave. 2022.
\newblock Atlas: Few-shot learning with retrieval augmented language models.
\newblock \emph{ArXiv e-prints}.

\bibitem[{Jiang et~al.(2025)Jiang, Liang, Zhu, Zhou, Ma, and Wang}]{jiang2025robustifyingvisionlanguagemodelsdynamic}
Tanqiu Jiang, Jiacheng Liang, Rongyi Zhu, Jiawei Zhou, Fenglong Ma, and Ting Wang. 2025.
\newblock \href {https://arxiv.org/abs/2505.17132} {Robustifying vision-language models via dynamic token reweighting}.
\newblock \emph{Preprint}, arXiv:2505.17132.

\bibitem[{Jiang et~al.(2024)Jiang, Wang, Liang, Li, Wang, and Wang}]{jiang2024robustkv}
Tanqiu Jiang, Zian Wang, Jiacheng Liang, Changjiang Li, Yuhui Wang, and Ting Wang. 2024.
\newblock Robustkv: Defending large language models against jailbreak attacks via kv eviction.
\newblock \emph{arXiv preprint arXiv:2410.19937}.

\bibitem[{Jovanovi{\'c} et~al.(2024)Jovanovi{\'c}, Staab, and Vechev}]{jovanovic2024watermark}
Nikola Jovanovi{\'c}, Robin Staab, and Martin Vechev. 2024.
\newblock Watermark stealing in large language models.
\newblock \emph{arXiv preprint arXiv:2402.19361}.

\bibitem[{{Kirchenbauer} et~al.(2023{\natexlab{a}}){Kirchenbauer}, {Geiping}, {Wen}, {Katz}, {Miers}, and {Goldstein}}]{john}
John {Kirchenbauer}, Jonas {Geiping}, Yuxin {Wen}, Jonathan {Katz}, Ian {Miers}, and Tom {Goldstein}. 2023{\natexlab{a}}.
\newblock A watermark for large language models.
\newblock In \emph{Proceedings of the IEEE Conference on Machine Learning (ICML)}.

\bibitem[{{Kirchenbauer} et~al.(2023{\natexlab{b}}){Kirchenbauer}, {Geiping}, {Wen}, {Shu}, {Saifullah}, {Kong}, {Fernando}, {Saha}, {Goldblum}, and {Goldstein}}]{para-attack}
John {Kirchenbauer}, Jonas {Geiping}, Yuxin {Wen}, Manli {Shu}, Khalid {Saifullah}, Kezhi {Kong}, Kasun {Fernando}, Aniruddha {Saha}, Micah {Goldblum}, and Tom {Goldstein}. 2023{\natexlab{b}}.
\newblock On the reliability of watermarks for large language models.
\newblock \emph{ArXiv e-prints}.

\bibitem[{{Krishna} et~al.(2023){Krishna}, {Song}, {Karpinska}, {Wieting}, and {Iyyer}}]{retrieval-defense}
Kalpesh {Krishna}, Yixiao {Song}, Marzena {Karpinska}, John {Wieting}, and Mohit {Iyyer}. 2023.
\newblock Paraphrasing evades detectors of ai-generated text, but retrieval is an effective defense.
\newblock In \emph{Proceedings of the Advances in Neural Information Processing Systems (NeurIPS)}.

\bibitem[{Kuditipudi et~al.(2023)Kuditipudi, Thickstun, Hashimoto, and Liang}]{rohith}
Rohith Kuditipudi, John Thickstun, Tatsunori Hashimoto, and Percy Liang. 2023.
\newblock Robust distortion-free watermarks for language models.
\newblock \emph{ArXiv e-prints}.

\bibitem[{Kudo(2018)}]{sentencepiece}
John Kudo, Takuand~Richardson. 2018.
\newblock {S}entence{P}iece: A simple and language independent subword tokenizer and detokenizer for neural text processing.
\newblock In \emph{Proceedings of the Conference on Empirical Methods in Natural Language Processing (EMNLP)}.

\bibitem[{Lee et~al.(2023)Lee, Hong, Ahn, Hong, Lee, Yun, Shin, and Kim}]{lee2023wrote}
Taehyun Lee, Seokhee Hong, Jaewoo Ahn, Ilgee Hong, Hwaran Lee, Sangdoo Yun, Jamin Shin, and Gunhee Kim. 2023.
\newblock Who wrote this code? watermarking for code generation.
\newblock \emph{arXiv preprint arXiv:2305.15060}.

\bibitem[{{Li} et~al.(2019){Li}, {Ji}, {Du}, {Li}, and {Wang}}]{text-bugger}
Jinfeng {Li}, Shouling {Ji}, Tianyu {Du}, Bo~{Li}, and Ting {Wang}. 2019.
\newblock Textbugger: Generating adversarial text against real-world applications.
\newblock In \emph{Proceedings of the Network and Distributed System Security Symposium (NDSS)}.

\bibitem[{Li()}]{law_stack_exchange}
Jonathan Li.
\newblock Law stack exchange dataset.
\newblock \url{https://huggingface.co/datasets/jonathanli/law-stack-exchange}.
\newblock Accessed: 2025-01-29.

\bibitem[{Liang et~al.(2025{\natexlab{a}})Liang, Jiang, Wang, Zhu, Ma, and Wang}]{liang2025autoranweaktostrongjailbreakinglarge}
Jiacheng Liang, Tanqiu Jiang, Yuhui Wang, Rongyi Zhu, Fenglong Ma, and Ting Wang. 2025{\natexlab{a}}.
\newblock \href {https://arxiv.org/abs/2505.10846} {Autoran: Weak-to-strong jailbreaking of large reasoning models}.
\newblock \emph{Preprint}, arXiv:2505.10846.

\bibitem[{Liang et~al.(2024)Liang, Pang, Li, and Wang}]{liang2024model}
Jiacheng Liang, Ren Pang, Changjiang Li, and Ting Wang. 2024.
\newblock Model extraction attacks revisited.
\newblock In \emph{Proceedings of the 19th ACM Asia Conference on Computer and Communications Security}, pages 1231--1245.

\bibitem[{Liang et~al.(2025{\natexlab{b}})Liang, Wang, Li, Zhu, Jiang, Gong, and Wang}]{liang2025graphrag}
Jiacheng Liang, Yuhui Wang, Changjiang Li, Rongyi Zhu, Tanqiu Jiang, Neil Gong, and Ting Wang. 2025{\natexlab{b}}.
\newblock Graphrag under fire.
\newblock \emph{arXiv preprint arXiv:2501.14050}.

\bibitem[{Liang et~al.(2023)Liang, Bommasani, Lee, Tsipras, Soylu, Yasunaga, Zhang, Narayanan, Wu, Kumar, Newman, Yuan, Yan, Zhang, Cosgrove, Manning, Re, Acosta-Navas, Hudson, Zelikman, Durmus, Ladhak, Rong, Ren, Yao, WANG, Santhanam, Orr, Zheng, Yuksekgonul, Suzgun, Kim, Guha, Chatterji, Khattab, Henderson, Huang, Chi, Xie, Santurkar, Ganguli, Hashimoto, Icard, Zhang, Chaudhary, Wang, Li, Mai, Zhang, and Koreeda}]{liang2022holistic}
Percy Liang, Rishi Bommasani, Tony Lee, Dimitris Tsipras, Dilara Soylu, Michihiro Yasunaga, Yian Zhang, Deepak Narayanan, Yuhuai Wu, Ananya Kumar, Benjamin Newman, Binhang Yuan, Bobby Yan, Ce~Zhang, Christian~Alexander Cosgrove, Christopher~D Manning, Christopher Re, Diana Acosta-Navas, Drew~Arad Hudson, Eric Zelikman, Esin Durmus, Faisal Ladhak, Frieda Rong, Hongyu Ren, Huaxiu Yao, Jue WANG, Keshav Santhanam, Laurel Orr, Lucia Zheng, Mert Yuksekgonul, Mirac Suzgun, Nathan Kim, Neel Guha, Niladri~S. Chatterji, Omar Khattab, Peter Henderson, Qian Huang, Ryan~Andrew Chi, Sang~Michael Xie, Shibani Santurkar, Surya Ganguli, Tatsunori Hashimoto, Thomas Icard, Tianyi Zhang, Vishrav Chaudhary, William Wang, Xuechen Li, Yifan Mai, Yuhui Zhang, and Yuta Koreeda. 2023.
\newblock Holistic evaluation of language models.
\newblock \emph{Transactions on Machine Learning Research}.

\bibitem[{Liu et~al.(2023{\natexlab{a}})Liu, Pan, Hu, Li, Wen, King, and Yu}]{aiweia}
Aiwei Liu, Leyi Pan, Xuming Hu, Shu'ang Li, Lijie Wen, Irwin King, and Philip~S Yu. 2023{\natexlab{a}}.
\newblock An unforgeable publicly verifiable watermark for large language models.
\newblock In \emph{Proceedings of the International Conference on Learning Representations (ICLR)}.

\bibitem[{{Liu} et~al.(2024){Liu}, {Pan}, {Hu}, {Meng}, and {Wen}}]{aiweib}
Aiwei {Liu}, Leyi {Pan}, Xuming {Hu}, Shiao {Meng}, and Lijie {Wen}. 2024.
\newblock A semantic invariant robust watermark for large language models.
\newblock In \emph{Proceedings of the International Conference on Learning Representations (ICLR)}.

\bibitem[{Liu et~al.(2023{\natexlab{b}})Liu, Pan, Lu, Li, Hu, Wen, King, and Yu}]{liu2023survey}
Aiwei Liu, Leyi Pan, Yijian Lu, Jingjing Li, Xuming Hu, Lijie Wen, Irwin King, and Philip~S Yu. 2023{\natexlab{b}}.
\newblock A survey of text watermarking in the era of large language models.
\newblock \emph{arXiv preprint arXiv:2312.07913}.

\bibitem[{Liu et~al.(2024{\natexlab{a}})Liu, Tang, Chen, Liu, Indu, Zou, Dai, Galan, Porter, Jia, Zhang, and Xiong}]{liu2024sequentialllmframeworkfashion}
Han Liu, Xianfeng Tang, Tianlang Chen, Jiapeng Liu, Indu Indu, Henry~Peng Zou, Peng Dai, Roberto~Fernandez Galan, Michael~D Porter, Dongmei Jia, Ning Zhang, and Lian Xiong. 2024{\natexlab{a}}.
\newblock \href {https://arxiv.org/abs/2410.11327} {Sequential llm framework for fashion recommendation}.
\newblock \emph{Preprint}, arXiv:2410.11327.

\bibitem[{Liu et~al.(2025{\natexlab{a}})Liu, Wen, Nair, Liu, Lou, Zhang, Yeoh, Vorobeychik, and Zhang}]{liu2025ecoloracommunicationefficientfederatedfinetuning}
Han Liu, Ruoyao Wen, Srijith Nair, Jia Liu, Wenjing Lou, Chongjie Zhang, William Yeoh, Yevgeniy Vorobeychik, and Ning Zhang. 2025{\natexlab{a}}.
\newblock \href {https://arxiv.org/abs/2506.02001} {Ecolora: Communication-efficient federated fine-tuning of large language models}.
\newblock \emph{Preprint}, arXiv:2506.02001.

\bibitem[{Liu et~al.(2024{\natexlab{b}})Liu, Wu, Yu, and Zhang}]{10646875}
Han Liu, Yuhao Wu, Zhiyuan Yu, and Ning Zhang. 2024{\natexlab{b}}.
\newblock \href {https://doi.org/10.1109/SP54263.2024.00120} {Please tell me more: Privacy impact of explainability through the lens of membership inference attack}.
\newblock In \emph{2024 IEEE Symposium on Security and Privacy (SP)}, pages 4791--4809.

\bibitem[{Liu et~al.(2025{\natexlab{b}})Liu, Liang, Tang, Ye, Ma, and Xi}]{liu2025datadefenserolecuration}
Xiaoqun Liu, Jiacheng Liang, Luoxi Tang, Muchao Ye, Weicheng Ma, and Zhaohan Xi. 2025{\natexlab{b}}.
\newblock \href {https://arxiv.org/abs/2410.02220} {Data to defense: The role of curation in customizing llms against jailbreaking attacks}.
\newblock \emph{Preprint}, arXiv:2410.02220.

\bibitem[{Liu et~al.(2025{\natexlab{c}})Liu, Liang, Yan, Jang, Mao, Ye, Jia, and Xi}]{liu2025cyber}
Xiaoqun Liu, Jiacheng Liang, Qiben Yan, Jiyong Jang, Sicheng Mao, Muchao Ye, Jinyuan Jia, and Zhaohan Xi. 2025{\natexlab{c}}.
\newblock Cyber defense reinvented: Large language models as threat intelligence copilots.
\newblock \emph{arXiv preprint arXiv:2502.20791}.

\bibitem[{Liu et~al.(2024{\natexlab{c}})Liu, Liang, Ye, and Xi}]{liu2024robustifying}
Xiaoqun Liu, Jiacheng Liang, Muchao Ye, and Zhaohan Xi. 2024{\natexlab{c}}.
\newblock Robustifying safety-aligned large language models through clean data curation.
\newblock \emph{arXiv preprint arXiv:2405.19358}.

\bibitem[{Liu et~al.(2023{\natexlab{c}})Liu, Hu, Zhang, and Sun}]{liu2023watermarking}
Yixin Liu, Hongsheng Hu, Xuyun Zhang, and Lichao Sun. 2023{\natexlab{c}}.
\newblock Watermarking text data on large language models for dataset copyright protection.
\newblock \emph{arXiv preprint arXiv:2305.13257}.

\bibitem[{Lu et~al.(2024{\natexlab{a}})Lu, Wang, Bao, Wang, Li, Wu, Jiang, Xu, Wang, and Gao}]{lu2024all}
Lei Lu, Zhepeng Wang, Runxue Bao, Mengbing Wang, Fangyi Li, Yawen Wu, Weiwen Jiang, Jie Xu, Yanzhi Wang, and Shangqian Gao. 2024{\natexlab{a}}.
\newblock All-in-one tuning and structural pruning for domain-specific llms.
\newblock \emph{arXiv preprint arXiv:2412.14426}.

\bibitem[{Lu et~al.(2024{\natexlab{b}})Lu, Liu, Yu, Li, and King}]{lu2024entropy}
Yijian Lu, Aiwei Liu, Dianzhi Yu, Jingjing Li, and Irwin King. 2024{\natexlab{b}}.
\newblock An entropy-based text watermarking detection method.
\newblock \emph{arXiv preprint arXiv:2403.13485}.

\bibitem[{{Lukas} et~al.(2021){Lukas}, {Jiang}, {Li}, and {Kerschbaum}}]{sok-image}
Nils {Lukas}, Edward {Jiang}, Xinda {Li}, and Florian {Kerschbaum}. 2021.
\newblock Sok: How robust is image classification deep neural network watermarking?
\newblock In \emph{Proceedings of the IEEE Symposium on Security and Privacy (S\&P)}.

\bibitem[{Lütkebohle()}]{gptzero}
Ingo Lütkebohle.
\newblock Gptzero.
\newblock \url{https://gptzero.me/}.

\bibitem[{Ma(2024)}]{ma2024towards}
Xiaolong Ma. 2024.
\newblock \emph{Towards Efficient AI for Science in Scalable and High Performance Distributed System}.
\newblock Ph.D. thesis, University of Nevada, Reno.

\bibitem[{Ma et~al.(2024)Ma, Yan, Yang, Foster, Papka, Liu, and Kettimuthu}]{10.1145/3629526.3645035}
Xiaolong Ma, Feng Yan, Lei Yang, Ian Foster, Michael~E. Papka, Zhengchun Liu, and Rajkumar Kettimuthu. 2024.
\newblock \href {https://doi.org/10.1145/3629526.3645035} {Malletrain: Deep neural networks training on unfillable supercomputer nodes}.
\newblock In \emph{Proceedings of the 15th ACM/SPEC International Conference on Performance Engineering}, ICPE '24, page 190–200, New York, NY, USA. Association for Computing Machinery.

\bibitem[{Miller(1995)}]{miller-1994-wordnet}
George~A. Miller. 1995.
\newblock Wordnet: a lexical database for english.
\newblock \emph{Commun. ACM}, 38(11):39–41.

\bibitem[{Mitchell et~al.(2023)Mitchell, Lee, Khazatsky, Manning, and Finn}]{detect_gpt}
Eric Mitchell, Yoonho Lee, Alexander Khazatsky, Christopher~D. Manning, and Chelsea Finn. 2023.
\newblock Detectgpt: Zero-shot machine-generated text detection using probability curvature.

\bibitem[{Munyer and Zhong(2023)}]{munyer2023deeptextmark}
Travis Munyer and Xin Zhong. 2023.
\newblock Deeptextmark: Deep learning based text watermarking for detection of large language model generated text.
\newblock \emph{arXiv preprint arXiv:2305.05773}.

\bibitem[{Openai()}]{chatgpt}
Openai.
\newblock Openai chatgpt blog.
\newblock \url{https://openai.com/blog/chatgpt}.

\bibitem[{Pan et~al.(2024)Pan, Liu, He, Gao, Zhao, Lu, Zhou, Liu, Hu, Wen et~al.}]{pan2024markllm}
Leyi Pan, Aiwei Liu, Zhiwei He, Zitian Gao, Xuandong Zhao, Yijian Lu, Binglin Zhou, Shuliang Liu, Xuming Hu, Lijie Wen, et~al. 2024.
\newblock Markllm: An open-source toolkit for llm watermarking.
\newblock \emph{arXiv preprint arXiv:2405.10051}.

\bibitem[{Papineni et~al.(2002)Papineni, Roukos, Ward, and Zhu}]{papineni2002bleu}
Kishore Papineni, Salim Roukos, Todd Ward, and Wei-Jing Zhu. 2002.
\newblock Bleu: a method for automatic evaluation of machine translation.
\newblock In \emph{Proceedings of the Annual Meeting of the Association for Computational Linguistics (ACL)}.

\bibitem[{{Piet} et~al.(2023){Piet}, {Sitawarin}, {Fang}, {Mu}, and {Wagner}}]{markmyword}
Julien {Piet}, Chawin {Sitawarin}, Vivian {Fang}, Norman {Mu}, and David {Wagner}. 2023.
\newblock Mark my words: Analyzing and evaluating language model watermarks.
\newblock \emph{ArXiv e-prints}.

\bibitem[{Pillutla et~al.(2021)Pillutla, Swayamdipta, Zellers, Thickstun, Welleck, Choi, and Harchaoui}]{pillutla2023mauve}
Krishna Pillutla, Swabha Swayamdipta, Rowan Zellers, John Thickstun, Sean Welleck, Yejin Choi, and Zaid Harchaoui. 2021.
\newblock {MAUVE}: Measuring the gap between neural text and human text using divergence frontiers.
\newblock In \emph{Proceedings of the Advances in Neural Information Processing Systems (NeurIPS)}.

\bibitem[{Ren et~al.(2023)Ren, Xu, Liu, Cui, Wang, Yin, and Tang}]{ren2023robust}
Jie Ren, Han Xu, Yiding Liu, Yingqian Cui, Shuaiqiang Wang, Dawei Yin, and Jiliang Tang. 2023.
\newblock A robust semantics-based watermark for large language model against paraphrasing.
\newblock \emph{arXiv preprint arXiv:2311.08721}.

\bibitem[{{Sankar Sadasivan} et~al.(2023){Sankar Sadasivan}, {Kumar}, {Balasubramanian}, {Wang}, and {Feizi}}]{umd-attack}
Vinu {Sankar Sadasivan}, Aounon {Kumar}, Sriram {Balasubramanian}, Wenxiao {Wang}, and Soheil {Feizi}. 2023.
\newblock Can ai-generated text be reliably detected?
\newblock \emph{ArXiv e-prints}.

\bibitem[{Sato et~al.(2023)Sato, Takezawa, Bao, Niwa, and Yamada}]{sato2023embarrassingly}
Ryoma Sato, Yuki Takezawa, Han Bao, Kenta Niwa, and Makoto Yamada. 2023.
\newblock Embarrassingly simple text watermarks.
\newblock \emph{arXiv preprint arXiv:2310.08920}.

\bibitem[{Singh and Zou(2023)}]{singh2023new}
Karanpartap Singh and James Zou. 2023.
\newblock New evaluation metrics capture quality degradation due to llm watermarking.
\newblock \emph{arXiv preprint arXiv:2312.02382}.

\bibitem[{Sun et~al.(2023)Sun, Du, Song, and Li}]{sun2023codemark}
Zhensu Sun, Xiaoning Du, Fu~Song, and Li~Li. 2023.
\newblock Codemark: Imperceptible watermarking for code datasets against neural code completion models.
\newblock In \emph{Proceedings of the 31st ACM Joint European Software Engineering Conference and Symposium on the Foundations of Software Engineering}, pages 1561--1572.

\bibitem[{Sun et~al.(2022)Sun, Du, Song, Ni, and Li}]{sun2022coprotector}
Zhensu Sun, Xiaoning Du, Fu~Song, Mingze Ni, and Li~Li. 2022.
\newblock Coprotector: Protect open-source code against unauthorized training usage with data poisoning.
\newblock In \emph{Proceedings of the ACM Web Conference 2022}, pages 652--660.

\bibitem[{Tang et~al.(2023)Tang, Feng, Liu, Yang, and Hu}]{tang2023did}
Ruixiang Tang, Qizhang Feng, Ninghao Liu, Fan Yang, and Xia Hu. 2023.
\newblock Did you train on my dataset? towards public dataset protection with cleanlabel backdoor watermarking.
\newblock \emph{ACM SIGKDD Explorations Newsletter}, 25(1):43--53.

\bibitem[{{Touvron} et~al.(2023){Touvron}, {Martin}, {Stone}, {Albert}, {Almahairi}, {Babaei}, {Bashlykov}, {Batra}, {Bhargava}, {Bhosale}, {Bikel}, {Blecher}, {Canton Ferrer}, {Chen}, {Cucurull}, {Esiobu}, {Fernandes}, {Fu}, {Fu}, {Fuller}, {Gao}, {Goswami}, {Goyal}, {Hartshorn}, {Hosseini}, {Hou}, {Inan}, {Kardas}, {Kerkez}, {Khabsa}, {Kloumann}, {Korenev}, {Singh Koura}, {Lachaux}, {Lavril}, {Lee}, {Liskovich}, {Lu}, {Mao}, {Martinet}, {Mihaylov}, {Mishra}, {Molybog}, {Nie}, {Poulton}, {Reizenstein}, {Rungta}, {Saladi}, {Schelten}, {Silva}, {Smith}, {Subramanian}, {Tan}, {Tang}, {Taylor}, {Williams}, {Kuan}, {Xu}, {Yan}, {Zarov}, {Zhang}, {Fan}, {Kambadur}, {Narang}, {Rodriguez}, {Stojnic}, {Edunov}, and {Scialom}}]{llama}
Hugo {Touvron}, Louis {Martin}, Kevin {Stone}, Peter {Albert}, Amjad {Almahairi}, Yasmine {Babaei}, Nikolay {Bashlykov}, Soumya {Batra}, Prajjwal {Bhargava}, Shruti {Bhosale}, Dan {Bikel}, Lukas {Blecher}, Cristian {Canton Ferrer}, Moya {Chen}, Guillem {Cucurull}, David {Esiobu}, Jude {Fernandes}, Jeremy {Fu}, Wenyin {Fu}, Brian {Fuller}, Cynthia {Gao}, Vedanuj {Goswami}, Naman {Goyal}, Anthony {Hartshorn}, Saghar {Hosseini}, Rui {Hou}, Hakan {Inan}, Marcin {Kardas}, Viktor {Kerkez}, Madian {Khabsa}, Isabel {Kloumann}, Artem {Korenev}, Punit {Singh Koura}, Marie-Anne {Lachaux}, Thibaut {Lavril}, Jenya {Lee}, Diana {Liskovich}, Yinghai {Lu}, Yuning {Mao}, Xavier {Martinet}, Todor {Mihaylov}, Pushkar {Mishra}, Igor {Molybog}, Yixin {Nie}, Andrew {Poulton}, Jeremy {Reizenstein}, Rashi {Rungta}, Kalyan {Saladi}, Alan {Schelten}, Ruan {Silva}, Eric~Michael {Smith}, Ranjan {Subramanian}, Xiaoqing~Ellen {Tan}, Binh {Tang}, Ross {Taylor}, Adina {Williams}, Jian~Xiang {Kuan}, Puxin {Xu}, Zheng {Yan}, Iliyan {Zarov},
  Yuchen {Zhang}, Angela {Fan}, Melanie {Kambadur}, Sharan {Narang}, Aurelien {Rodriguez}, Robert {Stojnic}, Sergey {Edunov}, and Thomas {Scialom}. 2023.
\newblock Llama 2: Open foundation and fine-tuned chat models.
\newblock \emph{ArXiv e-prints}.

\bibitem[{Tu et~al.(2023)Tu, Sun, Bai, Yu, Hou, and Li}]{waterbench}
Shangqing Tu, Yuliang Sun, Yushi Bai, Jifan Yu, Lei Hou, and Juanzi Li. 2023.
\newblock Waterbench: Towards holistic evaluation of watermarks for large language models.
\newblock \emph{arXiv preprint arXiv:2311.07138}.

\bibitem[{Wang et~al.(2024{\natexlab{a}})Wang, Wang, and Xian}]{wangtim2024}
Di~Wang, Yuhui Wang, and Xiaochen Xian. 2024{\natexlab{a}}.
\newblock \href {https://doi.org/10.1109/TIM.2024.3374288} {A latent variable-based multitask learning approach for degradation modeling of machines with dependency and heterogeneity}.
\newblock \emph{IEEE Transactions on Instrumentation and Measurement}, 73:1--15.

\bibitem[{Wang et~al.(2024{\natexlab{b}})Wang, Yang, Chen, Zhou, Lin, Meng, Zhou, and Sun}]{lean}
Lean Wang, Wenkai Yang, Deli Chen, Hao Zhou, Yankai Lin, Fandong Meng, Jie Zhou, and Xu~Sun. 2024{\natexlab{b}}.
\newblock Towards codable text watermarking for large language models.
\newblock In \emph{Proceedings of the International Conference on Learning Representations (ICLR)}.

\bibitem[{Wang et~al.(2025{\natexlab{a}})Wang, Wang, Wang, and Wang}]{wangtase2025}
Yuhui Wang, Andi Wang, Di~Wang, and Dong Wang. 2025{\natexlab{a}}.
\newblock \href {https://doi.org/10.1109/TASE.2024.3504595} {Deep learning-based sensor selection for failure mode recognition and prognostics under time-varying operating conditions}.
\newblock \emph{IEEE Transactions on Automation Science and Engineering}, 22:9252--9274.

\bibitem[{Wang and Wang(2023)}]{wangcase2023}
Yuhui Wang and Di~Wang. 2023.
\newblock \href {https://doi.org/10.1109/CASE56687.2023.10260339} {An entropy- and attention-based feature extraction and selection network for multi-target coupling scenarios}.
\newblock In \emph{2023 IEEE 19th International Conference on Automation Science and Engineering (CASE)}, pages 1--6.

\bibitem[{Wang et~al.(2025{\natexlab{b}})Wang, Zhu, and Wang}]{wang2025selfdestructivelanguagemodel}
Yuhui Wang, Rongyi Zhu, and Ting Wang. 2025{\natexlab{b}}.
\newblock \href {https://arxiv.org/abs/2505.12186} {Self-destructive language model}.
\newblock \emph{Preprint}, arXiv:2505.12186.

\bibitem[{Wang et~al.(2024{\natexlab{c}})Wang, Bao, Wu, Liu, Yang, Zhan, Zheng, Jiang, and Zhang}]{wang2024self}
Zhepeng Wang, Runxue Bao, Yawen Wu, Guodong Liu, Lei Yang, Liang Zhan, Feng Zheng, Weiwen Jiang, and Yanfu Zhang. 2024{\natexlab{c}}.
\newblock Self-guided knowledge-injected graph neural network for alzheimer’s diseases.
\newblock In \emph{International Conference on Medical Image Computing and Computer-Assisted Intervention}, pages 378--388. Springer.

\bibitem[{Wang et~al.(2024{\natexlab{d}})Wang, Bao, Wu, Taylor, Xiao, Zheng, Jiang, Gao, and Zhang}]{wang2024unlocking}
Zhepeng Wang, Runxue Bao, Yawen Wu, Jackson Taylor, Cao Xiao, Feng Zheng, Weiwen Jiang, Shangqian Gao, and Yanfu Zhang. 2024{\natexlab{d}}.
\newblock Unlocking memorization in large language models with dynamic soft prompting.
\newblock In \emph{Proceedings of the 2024 Conference on Empirical Methods in Natural Language Processing}, pages 9782--9796.

\bibitem[{Wieting et~al.(2022)Wieting, Gimpel, Neubig, and Berg-kirkpatrick}]{wieting2021paraphrastic}
Johnand Wieting, Kevinand Gimpel, Grahamand Neubig, and Taylor Berg-kirkpatrick. 2022.
\newblock Paraphrastic representations at scale.
\newblock In \emph{Proceedings of the Conference on Empirical Methods in Natural Language Processing (EMNLP)}.

\bibitem[{Wu et~al.(2020{\natexlab{a}})Wu, Wang, Jia, Shi, and Hu}]{wu2020intermittent}
Yawen Wu, Zhepeng Wang, Zhenge Jia, Yiyu Shi, and Jingtong Hu. 2020{\natexlab{a}}.
\newblock Intermittent inference with nonuniformly compressed multi-exit neural network for energy harvesting powered devices.
\newblock In \emph{2020 57th ACM/IEEE Design Automation Conference (DAC)}, pages 1--6. IEEE.

\bibitem[{Wu et~al.(2020{\natexlab{b}})Wu, Wang, Shi, and Hu}]{wu2020enabling}
Yawen Wu, Zhepeng Wang, Yiyu Shi, and Jingtong Hu. 2020{\natexlab{b}}.
\newblock Enabling on-device cnn training by self-supervised instance filtering and error map pruning.
\newblock \emph{IEEE Transactions on Computer-Aided Design of Integrated Circuits and Systems}, 39(11):3445--3457.

\bibitem[{Wu et~al.(2024)Wu, Hu, Guo, Zhang, and Huang}]{dipmark}
Yihan Wu, Zhengmian Hu, Junfeng Guo, Hongyang Zhang, and Heng Huang. 2024.
\newblock A resilient and accessible distribution-preserving watermark for large language models.
\newblock In \emph{Forty-first International Conference on Machine Learning}.

\bibitem[{Xu et~al.(2024)Xu, Yao, and Liu}]{xu2024learning}
Xiaojun Xu, Yuanshun Yao, and Yang Liu. 2024.
\newblock Learning to watermark llm-generated text via reinforcement learning.
\newblock \emph{arXiv preprint arXiv:2403.10553}.

\bibitem[{Yang et~al.(2023)Yang, Chen, Zhang, Liu, Qi, Zhang, Fang, and Yu}]{yang2023watermarking}
Xi~Yang, Kejiang Chen, Weiming Zhang, Chang Liu, Yuang Qi, Jie Zhang, Han Fang, and Nenghai Yu. 2023.
\newblock Watermarking text generated by black-box language models.
\newblock \emph{arXiv preprint arXiv:2305.08883}.

\bibitem[{Yin et~al.(2025)Yin, Wang, Xu, Zhuang, Mozumder, Smith, and Zhang}]{yin2025digitalforensicsagelarge}
Zhipeng Yin, Zichong Wang, Weifeng Xu, Jun Zhuang, Pallab Mozumder, Antoinette Smith, and Wenbin Zhang. 2025.
\newblock \href {https://arxiv.org/abs/2504.02963} {Digital forensics in the age of large language models}.
\newblock \emph{Preprint}, arXiv:2504.02963.

\bibitem[{Yoo et~al.(2023)Yoo, Ahn, and Kwak}]{yoo2023advancing}
KiYoon Yoo, Wonhyuk Ahn, and Nojun Kwak. 2023.
\newblock Advancing beyond identification: Multi-bit watermark for language models.
\newblock \emph{arXiv preprint arXiv:2308.00221}.

\bibitem[{Yoo et~al.(2024)Yoo, Ahn, and Kwak}]{kiyoon}
KiYoon Yoo, Wonhyuk Ahn, and Nojun Kwak. 2024.
\newblock Advancing beyond identification: Multi-bit watermark for language models.
\newblock In \emph{Proceedings of the Annual Conference of the North American Chapter of the Association for Computational Linguistics (NAACL)}.

\bibitem[{Zawad et~al.(2025)Zawad, Ma, Yi, Li, Zhang, Yang, Yan, and He}]{zawad2025fedcust}
Syed Zawad, Xiaolong Ma, Jun Yi, Cheng Li, Minjia Zhang, Lei Yang, Feng Yan, and Yuxiong He. 2025.
\newblock Fedcust: Offloading hyperparameter customization for federated learning.
\newblock \emph{Performance Evaluation}, 167:102450.

\bibitem[{{Zhang} et~al.(2023){Zhang}, {Edelman}, {Francati}, {Venturi}, {Ateniese}, and {Barak}}]{watermark-sand}
Hanlin {Zhang}, Benjamin~L. {Edelman}, Danilo {Francati}, Daniele {Venturi}, Giuseppe {Ateniese}, and Boaz {Barak}. 2023.
\newblock Watermarks in the sand: Impossibility of strong watermarking for generative models.
\newblock \emph{ArXiv e-prints}.

\bibitem[{Zhang et~al.(2023)Zhang, Hussain, Neekhara, and Koushanfar}]{remarkllm}
Ruisi Zhang, Shehzeen~Samarah Hussain, Paarth Neekhara, and Farinaz Koushanfar. 2023.
\newblock Remark-llm: A robust and efficient watermarking framework for generative large language models.
\newblock \emph{ArXiv e-prints}.

\bibitem[{Zhang et~al.(2020)Zhang, Kishore, Wu, Weinberger, and Artzi}]{zhang2020bertscore}
Tianyi Zhang, Varsha Kishore, Felix Wu, Kilian~Q. Weinberger, and Yoav Artzi. 2020.
\newblock Bertscore: Evaluating text generation with bert.
\newblock In \emph{Proceedings of the International Conference on Learning Representations (ICLR)}.

\bibitem[{Zhao et~al.(2023)Zhao, Ananth, Li, and Wang}]{xuandong}
Xuandong Zhao, Prabhanjan Ananth, Lei Li, and Yu-Xiang Wang. 2023.
\newblock Provable robust watermarking for ai-generated text.
\newblock \emph{ArXiv e-prints}.

\bibitem[{Zhao et~al.(2024)Zhao, Gunn, Christ, Fairoze, Fabrega, Carlini, Garg, Hong, Nasr, Tramer et~al.}]{zhao2024sok}
Xuandong Zhao, Sam Gunn, Miranda Christ, Jaiden Fairoze, Andres Fabrega, Nicholas Carlini, Sanjam Garg, Sanghyun Hong, Milad Nasr, Florian Tramer, et~al. 2024.
\newblock Sok: Watermarking for ai-generated content.
\newblock \emph{arXiv preprint arXiv:2411.18479}.

\bibitem[{Zhu et~al.(2025)Zhu, Wang, Jiang, Liang, and Wang}]{zhu2025selfimprovingmodelsteering}
Rongyi Zhu, Yuhui Wang, Tanqiu Jiang, Jiacheng Liang, and Ting Wang. 2025.
\newblock \href {https://arxiv.org/abs/2507.08967} {Self-improving model steering}.
\newblock \emph{Preprint}, arXiv:2507.08967.

\bibitem[{Zhuang and Guan(2025)}]{zhuang2025largelanguagemodelshelp}
Jun Zhuang and Chaowen Guan. 2025.
\newblock \href {https://arxiv.org/abs/2502.13166} {Large language models can help mitigate barren plateaus}.
\newblock \emph{Preprint}, arXiv:2502.13166.

\bibitem[{Zhuang et~al.(2025)Zhuang, Jin, Zhang, Kang, Zhang, Dagher, and Wang}]{zhuang2025exploring}
Jun Zhuang, Haibo Jin, Ye~Zhang, Zhengjian Kang, Wenbin Zhang, Gaby~G Dagher, and Haohan Wang. 2025.
\newblock Exploring the vulnerability of the content moderation guardrail in large language models via intent manipulation.
\newblock \emph{arXiv preprint arXiv:2505.18556}.

\end{thebibliography}
